# Online detection and sorting of extracellularly recorded action potentials in human medial temporal lobe recordings, in vivo

Ueli Rutishauser[1,4], Erin M. Schuman[4], Adam N. Mamelak[2,3]

1) Computation and Neural Systems, California Institute of Technology, Pasadena CA, 91125

2) Epilepsy and Brain Mapping Program, Huntington Memorial Hospital, Pasadena CA 91105

3) Maxine Dunitz Neurosurgical Institute, Cedars-Sinai Medical Center, Los Angeles, CA 90048

4) Howard Hughes Medical Institute and Division of Biology 114-96,
California Institute of Technology, Pasadena CA 91125

Number of figures: 9
Number of tables: 4

Corresponding Author:

Erin M. Schuman, HHMI, Division of Biology, California Institute of Technology, MC 114-96, Pasadena CA 91125. Tel 6263958390, Fax 6265680631, e-mail schumane@its.caltech.edu

**Abstract**

Understanding the function of complex cortical circuits requires the simultaneous recording of action potentials from many neurons in awake and behaving animals. Practically, this can be achieved by extracellularly recording from multiple brain sites using single wire electrodes. However, in densely packed neural structures such as the human hippocampus, a single electrode can record the activity of multiple neurons. Thus, analytic techniques that differentiate action potentials of different neurons are required. Offline spike sorting approaches are currently used to detect and sort action potentials after finishing the experiment. Because the opportunities to record from the human brain are relatively rare, it is desirable to analyze large numbers of simultaneous recordings quickly using online sorting and detection algorithms. In this way, the experiment can be optimized for the particular response properties of the recorded neurons. Here we present and evaluate a method that is capable of detecting and sorting extracellular single-wire recordings in realtime. We demonstrate the utility of the method by applying it to an extensive data set we acquired from chronically-implanted depth electrodes in the hippocampus of human epilepsy patients. This dataset is particularly challenging because it was recorded in a noisy clinical environment. This method will allow the development of "closed-loop" experiments, which immediately adapt the experimental stimuli and/or tasks to the neural response observed.

## Introduction

Recent technological advances have made it possible to simultaneously record the activity of large numbers of neurons in awake and behaving animals using implanted extracellular electrodes.  In densely packed neuronal structures such as the cortex and the hippocampus the activity of multiple neurons can be recorded from a single extracellular electrode.  A complete understanding of neural function requires knowledge of the activity of many single neurons and it is thus crucial to accurately attribute every single spike observed to a particular neuron.  This task is greatly complicated by uncertainties arising from noise caused by firing of nearby neurons, inherent variability of spike waveforms due to bursts or fast changes in ion channel activation/deactivation, uncontrollable movement of the electrodes as well as external electrical noise from the environment.

There are two different ways to acquire and analyze electrophysiological data: i) store the raw electrical potential observed on all electrodes and perform spike detecting and sorting later (*offline sorting*) or ii) detect and sort spikes immediately (during acquisition) and only store the sorted spikes (*realtime online sorting*).  A combination of the above approaches is to detect spikes online and only store the detected spikes for later offline sorting. While it is reasonable to use offline sorting methods in certain cases, it is becoming increasingly necessary to develop realtime online sorting methods.  There are three main reasons to use such methods: i) Realtime online decoding allows "closed-loop" experiments, e.g. the adaptation of the experiment to the specific neural responses observed (compare to dynamic clamp on the single cell level, e.g. see (Prinz et al. 2004) for a review), ii) Fast data analysis: sophisticated offline spike sorting methods require extensive amounts of computation whereas online sorting allows immediate data analysis, iii) massive reduction in data transmission and storage.  Moving from offline sorting to realtime online sorting requires two separate technological advances: i) developing an online spike detection and sorting algorithm and ii) developing a realtime implementation of this algorithm.  The first condition is strictly necessary before a realtime version can be implemented and presents the main methodological challenge that needs to be addressed.  An algorithm that is online only uses information available at the current point in time and not information available in the future.  Applied to our approach, “online sorting” means that a spike observed at time *t* is sorted only using all information observed prior to and including point of time *t*. This is in contrast to offline sorting algorithms, which require that all spikes are available before sorting can start and thus require that all data is acquired and stored beforehand. Removing this requirement for total spike availability presents a formidable challenge and we focus exclusively on doing so in this paper. Note that it will be possible to implement the algorithm presented here for realtime analysis of many channels in parallel; this will be the focus of our future efforts.

While the problem of offline sorting has been intensively investigated (for a review see (Lewicki 1998), but also see (Abeles and Goldstein 1977; Fee et al. 1996a; Harris et al. 2000; Pouzat et al. 2004; Pouzat et al. 2002; Quiroga et al. 2004; Redish 2003; Sahani et al. 1998; Shoham et al. 2003)), relatively little work has been done on online sorting. Early attempts at online sorting focused on techniques which require manual definition of each cluster before sorting commences (Nicolelis et al. 1997). Other online classification approaches require a learning phase, after which neurons are classified in realtime (Aksenova et al. 2003; Chandra and Optican 1997). The disadvantage of this class of online methods is that only neurons which fire during the learning phase can be classified. In addition, if the spike shapes change during the experiment, the neuron can no longer be recognized. In this paper, we present and demonstrate an online spike detection and sorting method. Spikes originating from different neurons are distinguished based on spike waveform shape and amplitude differences, features which are unique for individual neurons. The algorithm iteratively updates the model and assigns spikes to clusters. It thus does not require a separate learning phase and is capable of detecting new neurons during the experiment. This feature is particularly crucial for experiments with human subjects because firing is very sparse and the “optimal”stimuli for recorded neurons are often unknown. As a result, it is not possible to excite all neurons during a learning phase that precedes the experiment. We will discuss this point further at a later stage in the paper.

We demonstrate our method by applying it to data recorded from arrays of single-wire depth electrodes that are semi-chronically implanted in the medial temporal lobe of human epilepsy patients. This analysis is particularly challenging because the data were acquired in an electrically noisy clinical setting without the option of re-positioning the electrodes to optimize spike detection. As a result, the data are compromised by low signal-to-noise ratios (SNR) as well as non-stationarities in the noise levels. Additionally, electrodes are implanted in densely packed neuronal structures (for example, the hippocampus), which complicates separating single-unit activity. These neurons generally have very low basal firing rates and can respond very selectively to certain stimuli.

Our experimental setup allows us to conduct long-term recordings simultaneously with complex behavioral experiments which can only be done with awake behaving humans. In these experiments, fast data analysis is highly desirable. Our patients are extremely rare (< 15 a year) and our recording sessions are short (1 – 4 hours). Although we can record for 1-5 days, the same neuron cannot be obtained with any reliability on subsequent recording days. There is always a tradeoff between sorting quality and fast data analysis, but in this kind of experiments it is crucial to know as fast as possible to what a neuron responded, so that the experiment can be adapted immediately. One possible compromise to achieve this is to use a simple, but online, algorithm which is capable of detecting most neurons and correctly sorting their spikes. This approach is reasonable for recordings from chronically implanted arrays of electrodes that do not allow for the individual movement of the electrodes to optimize response properties. Additionally, implanted arrays allow the simultaneous recording of many neurons over a long period of time and thus yield large amounts of data. However, it has proven difficult to store, process and analyze these large data sets because efficient methods for processing and analysis are lacking (see (Buzsaki 2004) for a discussion of these issues). An online spike detection and sorting algorithm, such as the one described below, will enable experimenters to process complex and large amounts of data in an efficient and effective way.

# Methods

### *Glossary of mathematical symbols and notation*

| Symbol | Definition |
|---|---|
| $\vec{S}_i$ | The raw waveform of spike $i$ |
| $\vec{M}_j$ | Mean waveform of cluster $j$ |
| $\lvert\vec{M}_k\rvert$ | Number of spikes assigned to cluster $k$ |
| $m$ | Total number of mean waveforms |
| $C$ | Number of spikes used to calculate mean waveforms (last N spikes assigned to each cluster) |
| $T_S, T_M$ | Threshold for sorting (S) and merging (M) |
| $N$ | Number of datapoints of a single waveform |
| $\vec{D}$ | Vector of distances |
| $\vec{Z}$ | Matrix of noise traces (with N datapoints each, each row is a noise trace) |
| $\vec{C}$ | Noise covariance matrix (dimensions: NxN) |
| $\vec{P}_i$ | Prewhitenend raw waveform of spike $i$ |
| $d_S, d_M$ | Distance between 2 clusters for sorting (S) and merging (M) |
| $d$ | Distance between 2 clusters (projection test) |

All population measurements are specified as mean ± standard deviation.

The raw waveform of spike $i$ is referred to as $\vec{S}_i$. A waveform is a vector that consists of N=256 datapoints. For every spike $i$, $\vec{S}_i(l)$ refers to the amplitude of the waveform at the sampling point $l$ ($l$ can take any value between $1...N$). $T$ denotes the threshold and is always a scalar. $f(t)$ and $p(t)$ refer to the bandpass filtered raw signal amplitude and the local energy at time point $t$ respectively.

### *Filtering and spike detection*

Spikes are detected using threshold crossings of a local energy measurement $p(t)$ of the bandpass filtered signal (Bankman et al. 1993; Kim and Kim 2003), which allows more reliable spike detection than thresholding the raw signal (Appendix A). If $p(t)$ is locally bigger than five times the standard deviation of $p(t)$, (or an other factor, referred to below as the extraction threshold), a candidate spike is detected (Csicsvari et al. 1998). For each threshold crossing (Figure 1C,D), a sample of 2.5 ms (64 samples at a 25 kHz sampling rate) is extracted from the filtered signal. This sample is upsampled 4 times using interpolation (Bremaud 2002), that is, by transforming the sample to Fourier space using FFT and back with more data points. After upsampling, the spike is sampled at 100 kHz and consists of N= 256 data points, with the maximum realigned at position 95: $\arg\max_{l}(S_i(l)) = S_i(95)$. Upsampling eliminates the roughness in the waveform introduced by undersampling the signal and the high-pass filtering and also allows a more accurate determination of the real peak of the waveform. Note that the peak of the waveform is typically not measured accurately because it is only reached for a very short time and thus often falls between points of time at which the signal is sampled.

### *Distance between the waveforms of two spikes*

The estimation of the number of neurons present, as well as the assignment of each spike to a neuron, is based on a distance metric between two spikes (Appendix A). Based on this distance, a threshold is used to decide i) how many neurons are present and ii) to assign each spike uniquely to one neuron or to noise, if unsortable. A crucial element of this approach is the threshold, which is calculated from the noise properties of the signal (Appendix A) and is equal to the squared average standard deviation of the signal, calculated with a sliding window. The threshold is thus not a parameter as it is automatically defined by the noise properties of the recording channel and is equal to (in a theoretical sense) the minimal signal-to-noise ratio required to be able to distinguish two neurons. It is assumed that the background noise is additive (see results) and the presence of a spike does not influence the noise properties (Fee et al. 1996b). It can thus be assumed that the variance of the noise of all waveforms of the same neuron is approximately constant (Pouzat et al. 2002). One concern is that the estimation of the threshold is strictly valid only if it is independent of the number of neurons and their spiking frequency on a specific channel. It is worth noting, however, that even if there exist multiple neurons each with high spiking frequency, most data points of the raw signal will not belong to a spike (but see (Quiroga et al. 2004)). We are thus assuming that the variance of the raw signal is approximately independent of the number of neurons

(Fee et al. 1996b).

### *Online sorting*

Each newly detected spike is sorted as soon as it is detected (Figure 2). The raw waveform of a newly detected, as of yet unsorted spike, is used to calculate the distance to all already known mean waveforms (clusters). The spike is assigned to the existing cluster to which it has minimal distance if the distance is smaller than a threshold value. If the minimal distance is larger than the threshold, a new cluster is automatically created. Every time a spike is assigned to a cluster, the mean waveform of that cluster is updated by taking the mean of the last C spikes that were assigned to this cluster. This causes the mean waveforms of each cluster to change as well, which might result in two clusters which have mean waveforms whose distance is less than the threshold. In this case, the two clusters become indistinguishable and they are thus merged. The spikes assigned to both clusters will be assigned to the newly created cluster (see Appendix B for details of the algorithm). Note that not every cluster created in this manner will represent a single unit. In fact, many small clusters will be created which represent noise. These can easily be discarded by requiring a minimal number of spikes for a valid cluster. However, noise of a stereotypic shape will create large clusters; these are also discarded. See the section below on how to evaluate potential single-unit clusters below for a discussion of this issue.

### *Calculating the threshold*

There are two thresholds used in the algorithm: The threshold for considering a new spike part of an existing cluster $T_S$ and the threshold for considering two clusters apart $T_M$ . We considered two possible ways of estimating these two thresholds from the background noise of the raw signal. Common to both are that they are calculated automatically from the data.

The first (exact) approach is to pre-whiten the waveforms of detected spikes using the covariance matrix of the noise (see Appendix D). In this way, the datapoints of a given waveform can be considered uncorrelated and the noise is white and of standard deviation 1 in each dimension (by design). The summed squared residuals of the difference between two waveforms (Eq 3b) can thus be considered $\chi^2$ distributed with the number of degrees of freedom equal to the number of datapoints that constitute a waveform. The threshold of the distance calculated as such can be estimated from the $\chi^2$ distribution (Eq 5). The distance between the mean waveforms of two clusters can be calculated as the square root of the summed squared residuals, which is, by definition, the standard deviation multiplied by the number of datapoints. The threshold for merging can thus be set in terms of number of standard deviations by which clusters should be separated until they are considered equal. This procedure allows us to estimate the two thresholds $T_S$ and $T_M$ automatically by using the covariance of the noise. While this is the statistically optimal estimate of the thresholds, it requires an accurate estimate of the covariance. This turns out to be a non-trivial task for real data and its iterative computation is computationally expensive. Additionally, pre-whitening requires computation of the inverse of the covariance matrix. Unfortunately, the determinant of the covariance matrix is often small (close to singularity), which makes this operation numerically unstable in some situations. To circumvent this problem we also tested the algorithm by using an approximated version of the threshold which does not require pre-whitening of the waveforms. The approximated thresholds (both $T_S$ and $T_M$ ) are equal to the variance of the raw signal (Eq 4a). The distance between two waveforms, both for sorting and merging, is calculated as the sum of the squared residuals of the difference between two waveforms (Eq 3a). Here, the raw waveforms (after upsampling and re-alignment) are used. No pre-whitening is performed.

In the results section we present performance estimates for both the exact as well as the approximation method for estimating the threshold.

### *Simulation of synthetic data*

Simulated raw data traces were generated by using a database of 150 mean waveforms taken from well-separated neurons recorded in previous experiments. To generate random background noise, a large number of those waveforms were randomly selected, randomly scaled and added to the noise traces. Executing this procedure many times resulted in realistic background noise, as judged by comparing the raw signal, the filtered signal and its autocorrelation (Figure 3) to the real data. This random background noise trace can be arbitrarily rescaled to a pre-specified standard deviation to simulate different noise situations. Noise is scaled to a standard deviation of 0.05, 0.10, 0.15 and 0.20.

Identifiable neurons are added by simulating a number of neurons (between 3 and 5 in the following cases) with a renewal Poisson process with a refractory period of 3ms and a fixed firing rate between 1 and 10Hz (which corresponds to the typical firing rate of real neurons in our data). For each neuron, one pre-defined mean waveform was used. Mean waveforms were re-scaled such that they were bounded in the range [-1..1] (arbitrary units). By systematically varying the noise levels, signal-to-noise ratios (SNR) comparable to those observed in real data were simulated. We calculate the SNR ratio (Eq 6 in Appendix A) as the root mean square value of the mean waveform divided by the standard deviation of the noise (Bankman et al. 1993). The average SNR is calculated by averaging the SNR of each waveform. To aid comparison, this method of generating simulated raw data traces was intentionally chosen to be essentially the same as the one used by Quian Quiroga et al. 2004 (Quiroga et al. 2004).

### ***Extracellular recordings***

We use data recorded from human patients implanted with hybrid chronic depth electrodes to treat drug resistant epileptic seizures. The electrodes contain an inner bundle of eight 50 µm microwires that extend approximately 5mm beyond the tip of the depth electrode (Fried et al. 1999). The clinical reason for implanting electrodes is to record electrical activity during epileptic seizures to locate the anatomical locus of seizure onset.

Electrodes were surgically removed approximately 2-4 weeks after implantation. Recording sessions, each 1-2h long, started approximately 48 hours after electrode implantation and lasted up to 4 days. We recorded extracellularly from 3 macroelectrodes with a total of 24 single channels (each connected to a single wire). One wire of each macroelectrode (with low impedance) was used for local grounding. Electrodes were implanted in the amygdala and hippocampi of subjects and data was recorded while subjects performed visual psychophysical experiments, similar to those reported in (Kreiman et al. 2000), as well as other behavioral experiments such as navigating in a virtual world. Data were acquired continuously with a low pass cut-off of 9 kHz, sampled at 25 kHz and stored for later analysis. The gain of the amplifiers (Neuralynx Inc) was set individually on a case-by-case basis (based on electrode impedance and noise) in the range of 20000 to 50000, with an additional A/D gain of 4.

All subjects gave informed consent to participate in the research, and the research was approved by the Institutional Review Boards of both Huntington Memorial Hospital and the California Institute of Technology. The location of the implanted electrodes was solely determined by clinical requirements for locating the seizure onset and the research team had no influence on electrode placement. The exact location of the electrodes was determined from high resolution structural MRI images taken immediately before and after electrode implantation.

### *Criteria to identify clusters representing single-units*

A collection of spikes is well separated if the following criteria are met: i) less than a small (e.g. < 3.0 %) percentage of all spikes have an ISI of less than 3ms (refractory period), ii) the power spectrum is within ± 5 standard deviations in the range of 20...100Hz, excluding < 20Hz because of theta/gamma oscillations), does not go to zero for high frequencies (Poisson process); note that at low frequencies (< 40Hz), a dip is expected due to the refractory period (Franklin and Bair 1995; Gabbiani and Koch 1999).

### *Quality of separation evaluation criteria*

We use a statistical tool commonly called a *projection test* to quantify both the degree of overlap between the clusters and the goodness-of-fit to the theoretically expected distribution of spikes around the cluster center. In the context of spike sorting this test was originally proposed by Pouzat et al. (Pouzat et al. 2002). We only summarize the procedure here and mention some additional problems associated with it (also see discussion and Appendix D): The raw waveforms are first pre-whitened (e.g. decorrelated) using the known autocorrelation (Figure 3) of pure noise segments (where no spikes were detected). Mathematically, this implies that the noise must be of full bandwidth and the covariance matrix of the noise traces is thus invertible. However, this is not always the case. See appendix D for further discussion of this issue. After this step, each datapoint of the raw waveform is independent of all the others, with white noise of standard deviation 1. This is done for the waveform of each detected spike. Afterwards, each waveform (with N datapoints) can be regarded as one point in N dimensional space. The center of a cluster is represented by the point in N dimensional space that corresponds to the mean of all waveforms assigned to the cluster. Since the noise is white with a known standard deviation of 1, the theoretically expected distribution of spikes of the same cluster around this center is known (a multivariate gaussian with a standard deviation of 1).

For any pair of clusters found on a single wire, the projection test can be applied to quantify the overlap between the two clusters. This is done by projecting the difference of every spike and the center of the cluster it is assigned to (residuals) onto the vector that connects the two centers of the clusters. This results in two distributions of a single one-dimensional quantity, centered on the two centers (Figures 5D and 7D). The distance between these two centers can conveniently be used as a measure of separation. If the distance is too small, one or both of the clusters have to be discarded. If the goodness-of-fit of the two clusters to the expected distribution is reasonably good (see below), then the overlap can be estimated: a distance of >5 guarantees an overlap of less than 1%, a distance >3.2 an overlap less than 5% and a distance of >2.8 an overlap of less than 7.5%. Please see the discussion for an application to our data.

For any given pair of clusters, the theoretically expected distribution (normal with standard deviation  = 1) of the projected residuals can be compared against the empirically observed distribution. We use a $R^2$ goodness-of-fit between the empirically-estimated probability density function and the theoretically-expected probability density function to quantify this. Note that the empirically-estimated distribution of the same cluster can look different if compared to different (other) clusters since the residuals are a projection of the residuals onto the vector connection the two centers (e.g. Figure 5D the first 2 subplots, where cluster 1 is compared against cluster 2 and 3). The projection test can either be applied post-hoc after sorting is finished or periodically (e.g. every few minutes) during the recording session. If it is applied periodically, clusters that don't qualify can be discarded automatically.

### *Implementation*

We implemented the proposed system in MATLAB (Mathworks, Natick MA) to assess its usefulness and evaluate its properties. The implementation is split into two parts: spike detection and sorting. Spike detection reads a raw data stream either from the network (broadcast by the acquisition system) or from a file and detects spikes. The raw data stream is in the Neuralynx (Neuralynx Inc, Tucscon AZ) NCS format. The detected spikes are passed on to the online sorting part, which sorts the spikes one-by-one, as they become available. The results of the sorting are stored and later analysed using the statistical methods described. Our implementation is not optimized for speed at this time. All running time measurements were made on the same machine (Intel Xeon 3Ghz) with MATLAB version R14SP1.

## Results and Discussion

### *Signal acquisition and filtering*

The continuously recorded signal (with a sampling rate of 25kHz, Figure 1A) is bandpass filtered by a 4-pole butterworth filter with a high-pass frequency of 300Hz and a low-pass cut off of 3000Hz (Figure 1B) to exclude both the low-frequency components, e.g. local field potentials (LFP), and high frequency components (noise) of the signal.

### *Spike detection*

Spike detection from raw data with high noise levels (Figure 1B) was reliably achieved using the local energy thresholding method (see Methods). Figure 1C demonstrates the advantage of the method: whereas the spikes between 8s and 10s (x axis) can not be detected in the filtered signal (Figure 1B), they are reliably picked up by the local energy signal (Figure 1C).

### *Waveform extraction and re-alignment*

For every spike detected, 64 data samples are extracted, with the peak at sample 25. The waveform is then upsampled 4x and re-aligned again, such that the peak is at sample 95 (see methods for details). Re-aligning twice, once before extraction and once after upsampling, is crucial because the upsampling will change the location of the peak.

The position of the peak is estimated more accurately after upsampling. Crucial is the accurate determination of where the peak of the waveform is located. This is, however, difficult and great care needs to be taken to avoid the erroneous splitting of one cluster into two because of re-alignment issues. This situation arises because we observe many very different waveforms in our recordings. Often the waveform has a dominant peak in either the positive or negative direction, but sometimes the situation is less obvious. Consider, for example, the 3 waveforms shown in Figure 4C. Whereas the blue and the red waveform have a dominant peak on the positive and negative side respectively, the situation for the green waveform is less clear. It has a peak of approximately the same amplitude in the negative and positive direction and either could be used for re-alignment. This situation is not artificial and arises often in our recordings (e.g. Figure 7A). If the simplest re-alignment procedure is chosen, e.g. re-align all spikes at their absolute maximal amplitude, the spikes originating from the green neuron shown would artificially be split into two clusters. This is because variance caused by noise would sometimes make the negative peak maximal and sometimes make the positive peak maximal. The strategy we have found to avoid this problem as best as possible is to use the order in which the peaks occur. If the peak in the negative direction appears before the peak in the positive direction, the waveform is re-aligned at the negative peak. If, on the other hand, the positive peak appears before the negative peak, the positive peak is used to re-align. Exceptions to this procedure are used if only one or none of the peaks are significant, that is, their peak amplitude is less than the standard deviation of the noise (see Algorithm 3 in Appendix C). Using this procedure, we can accurately re-align and sort spikes such as the one shown in Figure 4C. However, there are still situations in which this method is not able to correctly realign spikes. For example, if the waveform of a neuron has a first peak which is barely significant and a peak which is highly significant, the cluster will be artificially split. This will only be the case for neurons which are close to the distinguishable signal-to-noise level and in our experience this case is rather rare. But in the rare occurrence, this problem is detected by the projection test and this cluster is then discarded.

### *Evaluation of sorting - synthetic data*

We performed spike detection and online sorting on synthetic data to evaluate the online algorithm's performance. Data were simulated to resemble the real data as closely as possible. Specifically, we observe that the noise in our data is strongly autocorrelated (Figure 3) and thus we do not assume independent Gaussian noise. Rather, the noise itself likely consists of many randomly mixed waveforms of unidentifiable neurons. Identifiable neurons are simulated as independent Poisson renewal processes with a pre-set firing rate (see Methods). Every time the simulated Poisson neuron fires, it's waveform is added to the noise trace. The waveforms, both for the simulated background noise and the simulated neurons, are chosen such that they closely resemble waveforms we have observed in previous experiments.

Since the mean waveform is added to the already generated noise trace, the added waveform will be corrupted by the strongly correlated background noise. As Poisson neurons fire independently it is possible that there are overlapping spikes. Since the background noise and the neuronal firing are independent, it will be the case that some of the spikes will not be detectable and thus the number of sortable spikes could be less than the number of spikes originally inserted. In addition, for real datasets, low sample rates, compared to the frequency of spike waveforms, can cause problems in spike sorting due to misaligned peaks (the real peak was not sampled). We include this effect in our simulated data by originally simulating the data at 4 times the sampling rate (100 kHz) and then downsampling the data afterwards (to 25 kHz) before it is used for detection. This reproduces the misalignment of peak values that can be observed in real datasets.

We used the approximation method for estimating the thresholds for sorting and merging. See the next section for a performance comparison of the two methods (exact and approximate) of estimating the threshold.

**Simulated Dataset 1:** This dataset contains 3 neurons (Figure 4), each simulated by a renewal Poisson process with a refractory period of 3ms and a mean firing rate of 5, 7 and 4 Hz, respectively. To provide equal SNR ratios for all waveforms, the mean waveforms of the 3 neurons were rescaled so that their peak amplitude was 1 (Figure 4C). A 100s background noise trace was simulated as described (see Methods) and scaled so that it had a standard deviation of 0.05, 0.10, 0.15 or 0.20. Neuronal firing was simulated for 100s each and the point of time at which each neuron fired was stored. For each of the 4 noise levels, the noise trace is rescaled appropriately and then the mean waveforms of the neurons are added to the trace at the timepoints the Poisson neuron fired. Using this procedure, there will be 4 traces with different noise levels that contain exactly the same noise (same signal, but different amplitude) and exactly the same neuronal firing (In Figure 4A-B, the noise trace with added firing for noise level 0.20 is shown).

The simulated raw data traces were processed exactly as real data is processed (bandpass filter, spike detection, spike extraction, online sorting). The different noise levels (1, 2, 3, and 4) were processed and evaluated independently (Table 1). They correspond to an SNR of 6.7, 3.4, 2.2 and 1.2 respectively. No parameters were modified or specified manually except the extraction threshold (row Thr in Table 1). The results of the algorithm were evaluated independently for both detection and sorting.

To illustrate how to read the detailed results in Table 1, we consider the results of one particular noise level (level 3, noise standard deviation = 0.15, SNR of waveforms 3.4). Theoretically, there were 475, 718 and 383 spikes, respectively, generated by the 3 neurons. Of those, 97% were correctly detected (448, 701 and 377). This implies that 3% of the generated spikes were not detectable, either because they were corrupted by noise and hence failed to cross the threshold or they were inappropriately aligned. Of the 1526 correctly detected spikes, 1407 were correctly assigned to one of the 3 clusters. 46 spikes were incorrectly assigned to one of the 3 clusters (false positives (FP)). False positives can be either true spikes which are assigned to the wrong cluster (misses) or noise waveforms inappropriately detected as spikes and then assigned to one of the clusters. Both forms of FP are shown in the table. In this case, 119 spikes were misses. The number of misses plus the number of correctly assigned (TP) equals the number of detected spikes. The number of TP plus FP equals the number of spikes assigned to a cluster. TP and FP are specified as percent (%) of total number of spikes assigned to a cluster.

This dataset demonstrates that the algorithm is capable of correctly sorting 3 distinguishable neurons with equal SNR. Even in the worst case, where the SNR equals 1.2, 79% of all spikes could be detected correctly and 89% of all spikes assigned to one of the 3 clusters were assigned correctly. Figure 4D illustrates the result for all 4 levels of noise and also indicates for each noise level the variance of individual waveforms. Figure 4A and 4B show an extract of a raw data trace with the most difficult noise level (SNR=1.2). This is a situation we commonly observe in our real data (see Figure 9A).

The results of dataset 1 thus demonstrate the basic capabilities and limits of the algorithm and the parametric choices made. With the following two datasets we will address more specific elements of the algorithm: the limits of detectability (spike detection) and the limits of discriminability (spike sorting).

**Simulated Dataset 2 - Limits of detectability:** This second set of data addresses the limits of detectability, that is, under what conditions will the spiking of a neuron become undetectable due to background noise. To address this issue, a more realistic situation is simulated: we simulated 3 neurons with mean waveforms of different peak amplitude and thus different SNR. The 3 waveforms are illustrated in Figure 5A. All other conditions of the simulation were the same as in dataset 1. The average SNR of the 4 noise levels is 5.2, 2.6, 1.7 and 1.3. However, the SNRs of the individual waveforms are not equal and some will thus be harder to detect (see Table 2 for details). An additional difficulty presented by the 3 mean waveforms in Figure 5A is that they all have approximately equal peak amplitudes in the negative and positive direction. This makes it more difficult and sometimes ambiguous where a spike should be re-aligned.

The algorithm's performance on dataset 2 is shown in Table 2. Looking at the case of noise level 3, with mean waveform SNRs of 1.4, 1.4 and 2.3 (average 1.7), 56%, 56% and 98% of the spikes of each unit could be detected, respectively. Compared to noise level 2, this presents a substantial drop in the percent detected for the first two units. Further looking at noise level 4, where the SNR of the first 2 neurons drops to 1.1, only 21% and 15% of the spikes were detected. The limits of our spike detection and re-alignment technique are thus between an SNR of 1.1 and 1.4 for waveforms which are difficult to re-align. Detectability is limited because low SNR spikes do not cross the spike detection threshold or, if they do cross the threshold, they can not be correctly realigned and are discarded (see section on re-alignment). For waveforms (e.g. unit 3 in this dataset) that possess an easily detectable peak, a substantial number of spikes can be correctly detected and re-aligned at relatively low SNR values (e.g. 70% for an SNR of 1.7). The extraction threshold (column labeled Thr in Table 2) used for the 4th noise level was 4.5, which is a conservative value compared to the value of 4.0 used in dataset 1. This value was chosen to diminish the false positive rate. The choice of the extraction threshold is always a trade-off between missed detections and false detections, but as can be seen in this simulation, a value of 4.5 seems to provide a good balance between these two opposing factors.

**Simulated Dataset 3 - Limits of Discriminability**: This dataset combines the factors addressed by dataset 1 and 2 and adds difficulty by using 5 simulated neurons (Figure 5B), some of which have very similar waveforms (basically just scaled versions of each other). This will, at high noise levels, lead to merging of similar neurons because they can no longer be distinguished from one another. Additionally, all 5 neurons have similar firing rates (5, 7, 4, 6, and 9 Hz respectively). The detailed results are listed in Table 3. Figure 5C shows part of the raw data trace for all 4 noise levels.

Consider noise level 2, with an average SNR of 2.3 (individual SNRs of 2.1, 1.9, 1.4, 2.4, 3.9), detection as well as sorting of all 5 units works reliably: 89% of all spikes were correctly detected and 87% of all sorted spikes were assigned to the correct cluster. Noise level 3 has an average SNR of 1.6 (individual SNRs of 1.4, 1.3, 0.9, 1.6, 2.6). Unit 3 becomes very hard to detect in this scenario and thus only 8% of all unit 3's spikes were correctly be detected. However, due to additional difficulties presented by this waveform (red mean waveform in Figure 5B) in terms of re-alignment, none of them could be sorted. This is because both peaks of the mean waveform have an amplitude that is less than the noise standard deviation, and thus due to precautions taken in the realignment procedure the spikes have been discarded. Also, the false positive rate increased markedly, indicating that clusters started to merge. Units 1 and 5, for example, were partially merged with most of the spikes of unit 1 missclassified as belonging to unit 5. Note that the two waveforms are very similar to each other (Magenta and Blue waveforms in Figure 5B). This makes it hard to discriminate these two units at high noise levels. Figure 5C illustrates the difficulties of detecting units with small SNRs in high levels of noise. Shown is the same data segment (length 1s) for all 4 levels of noise.

The merging of neurons poses a unique problem- can we detect merging without knowing the true number of neurons (as is the case in real recordings)? To accomplish this, the projection test can be used. As illustrated in figure 5D, the projection test quantifies the overlap between every pair of clusters. For each cluster, the distribution of the residuals around the mean projected onto the line between the two mean waveforms in high dimensional space is shown. Due to transformations applied to the data to calculate this test (see methods), the residuals distribute (if sorting is perfect) around the mean with standard deviation = 1. This knowledge can be used to estimate two important factors: i) do spikes which were assigned to one cluster really belong to one cluster? and ii) are two clusters separate enough so as to be considered independent? The answer to the first question can be addressed by evaluating the goodness-of-fit of a normal distribution with standard deviation = 1. We use an $R^2$ value to do so. The closer to 1.0 this value is, the better is the fit. In case of corrupted clusters, the distribution will start to be skewed to one side and the $R^2$ value will be lower (for example, the combination 1->4 in figure 5D). The second question can be addressed by measuring the distance between two neurons (in terms of standard deviations). If two clusters are too close to each other to be accurately separated, they overlap (e.g. 1->5 and 3->4 in figure 5D, where the distance between the means is 4.6 and 5.0 standard deviations respectively). If both clusters that are compared are well fit by a normal distribution, a theoretical minimal distance can be calculated by setting an upper bound of overlap between the two normal distributions (e.g. Distance >= 5 equals less than 1% overlap).

### *Comparison between (exact and approximate) threshold calculation methods*

In the methods section, we compare two different ways of calculating the threshold: a computationally cheap method that approximates the threshold and a computationally more demanding method that calculates the statistically optimal threshold (see methods). In the previous section we used the approximation method to calculate the threshold. We repeated the same analysis for all 3 simulated datasets using the exact threshold calculation method. The results are illustrated in Table 4 and figure 6. The mean improvement in true positive rates for the 3 simulations is 2.9%, 3.1% and 2.6%. By definition, false positives are lowered by the same percentages. Also, in simulation 3 the exact threshold estimation method found 4 of the 5 existing clusters for the 2 most difficult noise levels. The exact threshold estimation method had its biggest advantage for the most difficult noise levels where it lead to an average true positive increase (and therefore false positives reduction) of 7.5%. On the other hand, the performance increase for the first 2 noise levels was only minor. It is thus only advantageous to use the exact estimation method if neurons are hard to distinguish and/or background noise is high.  In those cases the removal of correlations caused by the background noise results in a remarkable performance increase. The information contained in the background noise is thus useful for improving performance, as others have demonstrated before for offline sorting algorithms (Pouzat et al. 2002).

### *Comparison with offline sorting algorithms*

We used the same simulated datasets as described in the previous section to evaluate how the performance of our algorithm compares to other algorithms. We used two commonly used algorithms. Both algorithms are offline sorting algorithms, that is, they require all data to be available before sorting starts. The first algorithm (referred to as Offline 1) we compared against is the well known *KlustaKwik* clustering algorithm (Harris et al. 2000). We used the first 10 principal components, computed using PCA (Jolliffe 2002), as features. The minimum number of clusters was set to 3 and the maximum number clusters to 30.  Otherwise, all parameters were set to the default values. All parameters were the same for all simulations and noise levels. The second algorithm we compared against is the *WaveClus* algorithm developed by (Quiroga et al. 2004), referred to as "Offline 2".  This algorithm is particularly relevant for our comparison because it has been used to sort similar data to ours (Quiroga et al. 2005). Since this algorithm selects its own features (wavelets) directly from the data, we used the waveforms as input features. For both algorithms, we used the publicly available version of the code written by the authors. To exclude influences on sorting performance of different detection methods, we used our detection method to detect spikes. Spikes were upsampled and re-aligned before processing. Both algorithms thus had the exact same input data. The clusters generated by the two algorithms were manually matched to the clusters which originally generated the data. Clusters which do not exist in the original data (overclustering, noise) were assigned to noise.

The results of the comparison are summarized in figure 6 and table 4. The performance of a given algorithm can not be reduced to a single number because depending on the experimental situation, different criteria of performance are most crucial for the experimenter. To allow a fair comparison, we calculated 4 performance measurements: true positives (TP), false positives (FP), number clusters found and misses. We calculated the TP/FP in terms of the percentage of all spikes assigned to a given cluster that actually belong to this cluster (true positives, TP). The false positives (FP) are thus by definition the difference between the TP and 100%. Misses are in percent of all detected spikes which were missassigned. This includes spikes which were assigned to background noise. Overall, we find that all algorithms perform remarkably similar on all datasets. This is particularly true for the first two noise levels (Fig 6A-C, levels 1 and 2). Performance differences are larger for the more difficult noise levels 3 and 4. While all algorithms show a drop in performance for this two levels, the two offline algorithms identify fewer clusters than our online algorithm. This is because in the high noise situations, some of the clusters become very small and partially overlap with other clusters. The differences between these clusters cannot be resolved if correlations introduced by the background noise are not taken into account. This explains why in the case of noise level 4 in simulation 2 (Fig 6B, red line) the online algorithm using the exact threshold clearly has the best performance of all algorithms compared. Generally we observed that the offline algorithms appear to artificially merge clusters earlier than our algorithm. This causes an increase in the number of false positives, which then decreases the number of true positives. This does not imply that less spikes were correctly assigned but is a consequence of our definition of true positives, which we believe is the most relevant for experimental purposes. We also observed that the offline sorting algorithms generally tend to overcluster – that is, they generate ficticious clusters. As these artificial clusters also tend to be small, they typically do not violate the refractory period condition of no ISIs <3ms. One possibility to avoid this problem is to use the projection test as a post-hoc test after sorting with one of the offline sorting algorithms.

### *Evaluation of sorting - real data*

We chose 2 datasets from 2 different recording sessions to demonstrate the application of the algorithm to real datasets. In both sessions, we recorded from the right and left hippocampus (RH, LH) and either from the right or left amygdala (RA, LA). These two recording sessions were chosen because the first one represents an example with a high number of neurons per channel (on average, 3.7 ± 1.7 neurons per active channel, range 1-7) and the second a more typical case of fewer, but hard to distinguish, neurons (On average 2.0 ± 0.8 neurons per active channel, range 1-3). Using these two examples demonstrates that the algorithm works reliably in both cases.

Using our algorithm as described, with all parameters automatically estimated from the data and the extraction threshold set to 5 (see simulations for how to find this value), we found a total of 76 well-separated single neurons that pass all statistical tests and visual inspection. Figure 7 shows the result and the statistical criteria used for one particular channel (a single-wire, implanted in the RA). A total of 9096 raw waveforms were detected, 7237 (80%) of which were assigned to one of the 5 well- separated single units (1682, 3669, 210, 142 and 1534 for each cluster respectively). In figure 7A (from left to right), an overlay of all raw waveforms, the mean waveforms, and the decorrelated raw waveforms and means are shown. Each neuron is color-matched across the whole figure (1=cyan, 2=yellow, 3=green, 4=red, 5=blue). For the first two neurons detected, the raw waveforms, the interspike interval histogram (ISI), the powerspectrum of the ISI and the autocorrelation of the ISI are shown in Figures 7B and C (from left to right). The pertinent features for evaluation that are used are as follows: the fraction of ISIs shorter than 3ms (specified in % of all ISIs), the absence of peaks in the powerspectrum and an approximately zero autocorrelation for small (<3ms) timelags. We find that only the combination of all 3 criteria allow a sufficient classification of clusters as single-unit or not. We, for example, often observe clusters which have a perfect ISI (no <3ms) but with large peaks in the powerspectrum caused by noise (e.g., 60Hz and harmonics). Such clusters have to be discarded. Other indications of potential problems are an autocorrelation which does not return to 0 at long (>100ms) timelags.

Applying the above criteria allows us to identify all well-defined clusters that might represent single units, but it is not sufficient For example, special concern is warranted if two mean waveforms appear to be linearly scaled versions of each other, without any other distinguishing features(e.g. neuron 1 and 2 in Figure 7). In contrast, some neurons (e.g. neuron 4 and 5 in Figure 7) are very similar on some, but importantly not all, indices. Two waveforms that are linearly scaled versions of each other could be the result of spike height attenuation during a burst or electrode movement. The artificial splitting of a single unit into multiple clusters as well as erroneous merging of two single-units into one cluster can be detected using the projection test. There are two indicators of the projection test that can be used to assess splitting and merging: the distance between the two means of the clusters and the goodness-of-fit of the empirical to the theoretical distribution. If the distance between the two means is not sufficiently large (e.g. > 5 for less than 1% overlap) and/or the goodness-of-fit to the distribution is bad, one or both of the clusters has to be discarded. Figure 7D illustrates this method for the 4 pairs of neurons in which overlap might be suspected. As the left panel in Figure 7D shows, the distance between neuron 1 and 2 is sufficiently large (6.6) and the fit to the distributions is very good. In contrast, the fit of neuron 4 (3rd panel, red) is less good but still sufficient. Also, a few outliers can be identified which represent missalignments (far right of red distribution). Another reason for poorly separated single units is the merging of two clusters representing unique units. This can also be detected by the projection test. In this case, the distribution of spikes around the mean will be too broad (long, fat tails), which is an indication for merged clusters. Such clusters represent multi-unit activity and can be used as such in the further analysis. It is also helpful to look at a post-hoc PCA plot of the first 2 principal components (Figure 8). The principal components are computed from the raw, not pre-whitened, waveforms. The color is assigned by the clustering algorithm. In this plot it is also evident that cluster 1 and 2 are indeed separate. From the PCA plot it is less clear whether clusters 4 and 5 are indeed separate. Consultation of the projection test (Figure 7D) confirms that the clusters are separate but also indicates that there is some degree of overlap, as can also be seen in the PCA plot.

For comparison, we repeated the sorting of the same detected waveforms as shown in Figure 7 with the *WaveClus* offline sorting algorithm (see offline algorithm section for details). The algorithm identified a very similar number of spikes for each cluster (same order as above: 1529, 3452, 197, 113 and 1513). No other clusters were found except for the noise cluster. In total it assigned 75% of the total 9096 detected waveforms to one of the 5 clusters.

Population data for all 76 sorted neurons is shown in Figure 9. The average SNR of all mean waveforms, calculated by using the noise standard deviation for each channel, was 2.12±0.85 (Figure 9A). This measurement defines the SNR typically observed in experiments and thus serves as a guideline for the estimation and verification of parameters using the simulated data. A good general indicator of separation quality is the percent of ISIs which are shorter than 3ms (on average 0.21 ± 0.27%, Figure 9B). For all channels on which there was more than one neuron we calculated the distance between all pairs of neurons on each channel. The average distance was 12 ± 5 (Figure 9C).

### *Bursts*

The calculation of the threshold for sorting (minimum distance between clusters required) thus far only takes into account variance due to extracellular sources. However, the waveforms of a single neuron also vary due to intracellular reasons, mainly due to spikes which follow each other with an interspike interval of less than 100ms (Fee et al. 1996b; Harris et al. 2000; Quirk and Wilson 1999). This additional variance needs to be accounted for. As such, it is necessary to assume a slightly higher threshold than is estimated from the background noise. If it is known that the data which is sorted does not contain bursts, this correction does not need to be applied. A rough estimate whether there are bursts or not can be made by looking at a plot of the first two principal components of all detected raw waveforms. If there are distinct elongated clusters, bursting neurons are probably present and a correction needs to be applied.

The extracellular waveform during short ISIs is changed in a characteristic way. Most features of the spike remain the same, but the amplitude changes. That is, the waveform is linearly scaled. This will mainly affect the peak region of the spike. In our case, the peak region occupies approximately 0.5ms. The overshoot region will also be scaled, but the increase in variance due to this is minor because of its smaller amplitude relative to the spike peak. Peak spike amplitudes can be attenuated by up to 40% (Quirk and Wilson 1999). To account for this, the variance used to calculate the threshold has to be increased by 40% for the 0.5ms region of the peak region. See Equations 4b and 4c in Appendix A for the calculation, which results in a correction factor for the threshold of approximately 1.2. The fact that short ISIs cause scaling of the extracellular waveform also has important implications for the evaluation of the sorting results. Cases where two seemingly well-separated clusters have mean waveforms which appear to be linearly scaled versions of each other can be further evaluated manually.

### *Non-stationarities of noise levels*

Depending on the environment, the levels of background noise can change over time. Whereas this problem is manageable for recordings done in a controlled research environment, it is not possible to control external noise levels in clinical or other uncontrolled (e.g. behavioral studies) environments. The ability to dynamically adapt to non-stationary noise levels is thus crucial. We adapt to changing noise levels on two timescales: for fast, high-powered bursts of noise, we immediately stop extracting waveforms until the burst is over (usually far less than 200ms). To slowly changing levels of noise we adapt by calculating the threshold (which is calculated from the standard deviation of p(x), see methods) for spike extraction as a running average over a long time window (e.g., 1 minute).

### *Computation cost*

Our implementation (details in the methods) serves as a proof of principle and is not optimized for speed. We nevertheless report approximate running times for the different stages of the algorithm to enable a comparison against other algorithms, but it should be noted that careful optimization and more efficient implementation in a compilable programming language such as C++ will provide substantial improvements over the numbers reported here. We measured the running times while sorting a session consisting of 21 active channels, each recorded in parallel over a duration of 35min. Raw data was read from data files from the harddisk (one file per channel) A total of 143947 spikes were detected (average 6854±5234 spikes per channel). Detection took on average 194±13s per channel. This includes detection, extraction of pure noise sweeps, calculation of the noise autocorrelation and pre-whitening of each spike detected. Per channel approximately 100000 noise traces (40 per second) were extracted. Sorting took on average 18.24±13.9s per channel. Considering the number of spikes on each channel, this results in a sorting speed of 376 spikes/s. In total, this allows processing of a single channel at approximately 10 times the duration of data acquisition (on average 3.5 minutes for each channel). Optimizing this implementation will allow the realtime processing of many hundreds of channels in realtime.

### *Future improvements*

There are multiple ways in which the procedure presented here could be improved. One issue that is currently not addressed in our implementation[1] is overlapping spikes, which are caused by two nearby neurons firing in synchrony or by neurons firing closely together by chance. If two close-by neurons are synchronized such that they always fire together in a systematic and consistent way, the overlapping spike becomes detectable because a distinct cluster will be created. However, in the more common situation where spikes overlap in widely different situations, such spikes would be disregarded and classified as noise. It is imaginable to also test for linear combinations of mean waveforms to allow classification of such combined spike events. Indeed such an approach has been proposed (Atiya 1992; Takahashi et al. 2003).

[1] Our implementation as well as the simulated datasets are available at http://emslab.caltech.edu/software/spikesorter.html

The proposed algorithm has so far only been applied to the sorting of data from single wire electrodes but it would be straight-forward to extend its usage also to tetrode data (Harris et al. 2000). Instead of one mean waveform per identified source there would be four mean waveforms. This would further enhance performance and reliability while still using the same principle.

The re-alignment procedure we have described allows the accurate realignment of many difficult cases, but sometimes it still fails. Accurate realignment is necessary because our distance measurement for comparing two spikes requires that the two spikes are accurately realigned (at the same position). If this is not the case, the procedure fails. There are two possible improvements that could be made to remedy this situation. One would be to enhance the distance measurement so that it does not rely on realignment (e.g. re-positioning the two waveforms on a case-by-case basis for each distance measurement or using a translation invariant distance measurement). The second improvement could utilize a combined spatial and frequency space measurement, as has been proposed (Rinberg et al. 2003).

Our algorithm assigns each spike to one cluster only. This decision is taken at the point of time the spike is detected ("hard clustering"). An alternative approach would be to assign each spike a probability to which cluster it belongs and update this probability as the model (mean waveforms) change over time ("soft clustering"). While we have not taken this approach it is imaginable that it could be implemented in the framework we present here. Because we build and update our model iteratively over time, it is indeed possible that the model converges to the wrong solution. This is rather unlikely, though, because if a cluster slowly converges towards an other cluster, the two cluster centers eventually get too close and they are merged. However, merges are never reversed. If two cluster are very close by and are merged erroneously this situation will never be resolved.

Soft clustering could possibly deal with this situation.

### *Conclusions and Relevance*

Here, we propose a general online sorting algorithm and demonstrate and evaluate its sorting ability by applying it to a challenging dataset recorded in a clinical environment. There are a wide variety of applications made possible by online sorting which we are only starting to explore. The experimental approach taken in most animal single-unit recordings involves first the design of an experiment and then the search for neurons that respond appropriately to the experimental task. Obviously, this type of experimental design requires that electrodes can be moved freely by the experimenter; this is not possible in human studies. Of the many limitations posed by a clinical environment, the most constraining one is that chronically implanted electrodes are at a fixed position that can not be moved (Fried et al. 1999). Thus, only the neurons that can be recorded in the vicinity of the electrode can be analyzed. While it is still possible to design a static experiment and observe a neuronal response, it is the case that most neurons will not react in any systematic way to the stimuli presented. As one does not have access to the response properties of neurons during the experiment, these (non-stimulus-related) spike events are recorded and then during offline analysis discovered to be essentially useless. Electrodes in epilepsy surgery patients are implanted in higher-level brain structures such as the medial temporal lobe (MTL), including the hippocampus or the amygdala, and prefrontal cortex. Unlike the response properties of neurons in the primary sensory cortices, MTL neuron responses are multi-sensory and complex (Brown and Aggleton 2001), and hence possess less predictable response properties.

Thus, to make the most of the information obtainable with chronic implants in humans the traditional approach has to be reversed: the experiment needs to adapt itself to the neuronal response observed. Creating an adaptive experiment poses significant technological challenges which need to be addressed. The work presented in this paper is one of the main required techniques to be able to conduct adaptive experiments. Online sorting for the first time allows the experimenter to conduct real "closed-loop" experiments in awake behaving animals, similar to what is already possible with dynamic-clamp in single cell experiments (Prinz et al. 2004). Such experiments will be designed to immediately react to the neuronal response observed to a certain stimulus.

Additionally, online sorting is tremendously useful for conducting extracellular recordings in a noisy environment, like the hospital room. It is very hard and often impossible to judge manually (by visual inspection) whether the signals visible in the raw data trace are of sortable neurons or not. This can make the decision on which amplifier settings to use and from which electrodes to record arbitrary and often wrong. We, for example, often face the situation that there are more electrodes implanted than we can record from simultaneously. As such, we have to make an on-the-spot decision about which subset of electrodes to record from. Using offline data analysis, it sometimes becomes clear that the best available electrode was not chosen because it was not possible to identify the spikes by visual inspection alone. On the other hand, channels which look active and interesting often turn out to be corrupted by noise, so that they can't be used. Online spike sorting, implemented in realtime, will enable the experimenter to make the best informed choices about which electrodes to include during an experiment.

Another possible area of application is brain-machine interfaces. It has been demonstrated that it is possible to decode intended movements using chronically implanted electrodes in non-human primates using single-cell spike data from motor cortex (reviewed in (Mussa-Ivaldi and Miller 2003)) and higher cortical areas, e.g. (Musallam et al. 2004). Combined with the recent development of microdrive-driven chronically implanted arrays of electrodes this will ultimately allow online control of cortically-controlled neural prosthetics (Schwartz 2004). The algorithms for decoding intentions of movements (Chapin 2004) depend on the ability to simultaneously record the activity of many single neurons over a long time and it is thus crucial that spikes can be detected and sorted reliably in realtime. This presents a particular challenge in the uncontrolled and noisy environments in which such devices will have to function. Moving from the well controlled laboratory environment to a noisy real-world environment will increase the difficulty of spike detection and sorting tremendously. Our algorithm could be of use for such applications.

## Acknowledgements

We would like to acknowledge the support we have received from the Staff of the Epilepsy Department at the Huntington Memorial Hospital.  We thank Gilles Laurent, Bryan Smith and Shreesh Mysore for critically reading drafts of this manuscript.

## Grants

This research was supported by the Sloan-Swartz Center for theoretical neurobiology (U.R.), the Howard Hughes Medical Institute and the Gimbel Discovery Fund.

# Appendix A

### *Spike detection*

The local energy, or power, $p(t)$ (eq. 1a) of the signal is the running square root of the average power of the signal $f(t)$ using a window size of 1ms (n=20 samples at 25kHz sampling), the approximate duration of a spike (Bankman et al. 1993). $\bar{f}(t)$ is the running average, going back $n$ samples in time. $p(t)$ can be efficiently calculated for a signal of arbitrary length using a convolution kernel or a running window in online decoding.

$$P(t) = \left\{ \frac{1}{n} \sum_{i=1}^{n} \left( f(t-i) - \overline{f}(t) \right)^2 \right\}^{\frac{1}{2}} \qquad (1)$$

$$\overline{f}(t) = \frac{1}{n} \sum_{i=1}^{n} f(t-i) \qquad (2)$$

### *Distance between waveforms*

The distance between two spikes $\vec{S}_i$ and $\vec{S}_j$ is calculated as a residual-sum-of-squares (Eq. 3a) for the approximated threshold method. For the exact threshold estimation method, the same equation applies because the covariance matrix $\Sigma$ in Eq 3b is equal to $I$ for pre-whitened waveforms (by definition). Note that this distance is generally used to calculate the distance between a spike and a mean waveform of a neuron and not two spikes.

$$d_S(\vec{S}_i, \vec{S}_j) = \sum_{k=1}^{N} \left( S_i(k) - S_j(k) \right)^2 \qquad (3a)$$

$$d_S(\vec{P}_i, \vec{P}_j) = (\vec{P}_i - \vec{P}_j)\Sigma^{-1}(\vec{P}_i - \vec{P}_j)^T \qquad (3b)$$

Calculating the distance between the means of two clusters is achieved differently for the two methods of estimating the threshold: i) for the approximated threshold, $d_M = d_S$ and ii) for the exact threshold, $d_M = \sqrt{d_S}$ (equal to Eq 11 in the projection test).

### *Calculation of the threshold*

There are two thresholds which need to be calculated: $T_S$ (sorting) and $T_M$ (merging). In the case of the approximated threshold method, $T_S = T_M = T$, whereas $T$ is calculated as shown in Eq 4a. $\langle \sigma_r \rangle$ is the average standard deviation of the filtered signal f(x), calculated continuously with a long (e.g. 1 minute) sliding window. For efficiency reasons, the distance calculated in Eq 3a is not divided by N to normalize for the number of datapoints, but rather the threshold is multiplied by N in Eq 4a. This is mathematically equivalent, but Eq 3 can be calculated more efficiently in matrix notation in this form.

$$T = N \langle \sigma_r \rangle^2 \quad \text{(4a)}$$

In the case of the exact threshold estimation method, the two thresholds are calculated differently: Since $d_S$ is $\chi^2$ distributed (Johnson and Wichern 2002), the distance that includes all points belonging to the cluster with probability $1-\alpha$ can be calculated from the $\chi^2$ distribution (Eq 5). The threshold $d_M$ for merging is simply the number of standard deviations clusters need to be apart to be considered separate, which we assumed to be 3. $\alpha$ is typically set to 0.05 or 0.10 (5%,10%) and $p$ is the number of degrees of freedom (see text).

$$P[(\vec{P}_i - \vec{P}_j)\Sigma^{-1}(\vec{P}_i - \vec{P}_j)^T \le \chi_p{}^2(\alpha)] = 1-\alpha \qquad (5)$$

### *Correction factor for bursts*

The distance as calculated by Eq 4 does not take into account systematic variability of the waveform for reasons other than extracellular noise. To account for systematic waveform changes, particularly in spike amplitude , a correction factor is applied to increase T appropriately (Eq 4b).

$$T_C = cT \qquad \text{(4b)}$$

The correction factor c is calculated as following (here, N is assumed to be 256 datapoints): A burst is going to scale the peak region of the spike, which occupies approximately $B = 50$ datapoints (0.5ms). Correcting T for $B$ datapoints using a higher variance and leaving the other $N - B$ with the baseline variance is calculated using Eq 4c.

$$T_C = T(\frac{N - B}{N} + \frac{b_c B}{N}) \qquad (4c)$$

The correction factor $b_c$ specifies how much the variance is assumed to increase due to this. A conservative estimate is $b_c = 2$. Using above numbers, this results in a correction factor of 1.2 as is used throughout this paper. This correction factor is only applied if the threshold is calculated using the approximation method.

### *Signal-to-noise ratio*

The signal-to-noise ratio is calculated as the root-mean-square (rms) of a spike divided by the standard deviation (Bankman et al. 1993) of the raw data trace (Eq 6).

$$SNR = \frac{\left\| \vec{S}_i \right\|}{\sqrt{N\sigma^2}} . \qquad (6)$$

## Appendix B

### *Online sorting*

For each detected spike $\vec{S}_i$ the distance of $\vec{S}_i$ to all mean waveforms is calculated. Using algorithm 1, a spike is associated to cluster $j$ if it meets the following criteria: i) $d(\vec{S}_i, \vec{M}_j)$ is minimal compared to all other mean waveforms and ii) $\min(d(\vec{S}_i, \vec{M}_j)) < T$. If these conditions are met, Algorithm 2 is used to assign $\vec{S}_i$ to the existing cluster that meets the conditions. Also, the mean waveform of the cluster is updated using the last $C$ spikes that were associated to this cluster. This change could potentially create overlapping clusters (and will do so especially when not many spikes have been processed), which are automatically merged by Algorithm 2 (see below).

***Algorithm 1***

Task: Assign newly detected spike $\vec{S}_i$ to cluster or create new cluster if necessary.

1: $d_j = d_S(\vec{M}_j, \vec{S}_i)$ for $j = 1...m$ {distance to all known clusters}

2: **if** $\min(d_1, d_2, ..., d_m) \leq T_S$ **then**

3: assignSpike( $\vec{S}_i$ ) {call Algorithm 2}

4: **else**

5: $m \Leftarrow m + 1$

6: $\vec{M}_m \Leftarrow \vec{S}_i$

7: **end if**

***Algorithm 2***

Task: Assign spike $\vec{S}_i$ to cluster and merge clusters if necessary

1: $j \Leftarrow \arg\min(d_1, d_2, ..., d_m)$

2: assign $\vec{S}_i$ to cluster $j$

3: $\vec{M}_j \Leftarrow \langle \vec{S}_k \rangle$, for $k = |\vec{M}_j| - C...|\vec{M}_j|$ {update mean waveform as average of last C assigned spikes}

4: $\vec{D} = d_M(\vec{M}_j, \vec{M}_i)$, for $i = 1...j-1, j+1...m$ {distance of update mean waveform to all other mean waveforms}

5: **while** $\min(\vec{D}) < T_M$

6: $k \Leftarrow \arg\min(\vec{D})$

7: merge cluster $j$ with cluster $k$

remove cluster $k$

reassign all $\vec{S}_i$ assigned to cluster k to cluster $j$

$\vec{D} = d_M(\vec{M}_j, \vec{M}_j)$, for $j = 1...m$ {distance between all mean waveforms}

**end while**

# Appendix C

### ***Spike realignment***

### ***Algorithm 3***

Task: Decide where the peak of $\vec{S}_i$ is that is to be used for realignment.

sigLevel $\Leftarrow 2 * \langle \sigma_r \rangle$ {twice the std of the raw signal, see Eq4}

**if** $abs(\min(\vec{S}_i)) >= sigLevel$ **and** $abs(\max(\vec{S}_i)) >= sigLevel$ **then**

{Align according to temporal order of peaks}

**if** $find(\vec{S}_i == \max(\vec{S}_i)) < find(\vec{S}_i == \min(\vec{S}_i))$ **then**

peakInd = $find(\vec{S}_i == \max(\vec{S}_i))$ {realign at positive peak}

**else**

peakInd = $find(\vec{S}_i == \min(\vec{S}_i))$ {realign at negative peak}

**end if**

**else**

**if** $(abs(\min(\vec{S}_i)) >= sigLevel$ **and** $abs(\max(\vec{S}_i)) < sigLevel)$ **or** $(abs(\min(\vec{S}_i)) < sigLevel$ **and** $abs(\max(\vec{S}_i)) >= sigLevel)$ **then**

{only one peak is significant, realign at it}

**if** $abs(\max(\vec{S}_i)) > abs(\min(\vec{S}_i)$ **then**

peakInd = $find(\vec{S}_i == \max(\vec{S}_i))$

**else**

peakInd = $find(\vec{S}_i == \min(\vec{S}_i))$

**end if**

**else**

{This spike can't be re-aligned, discard}

**end if**

**end if**

# Appendix D

### ***Pre-whitening of waveforms***

The raw waveform, consisting of N datapoints, is corrupted by strongly correlated noise. To de-correlate the noise, that is, make each datapoint statistically independent of each other, a pre-whitening procedure (Kay 1993) is applied as following. A large number of noise traces (usually many 1000) is extracted from the same raw data signal as the spike waveforms but from the parts where no spike is detected. Each noise trace has the same number of datapoints as a spike waveform (N). Arranging all this traces in a matrix large matrix $\vec{Z}$ (each row is one noise trace), the covariance matrix $\vec{C}$ of the noise can be calculated (Eq 7). Using the Cholesky decomposition (Eq 8), this matrix can be decomposed such that the product of the resulting matrix multiplied by its inverse results in the original matrix $\vec{C}$ (Eq 9).

$$\vec{C} = \mathrm{cov}(\vec{Z}) \quad (7)$$

$$\vec{R} = chol(\vec{C}) \quad (8)$$

$$\vec{C} = \vec{R}'\vec{R} \quad (9)$$

By multiplying each raw spike waveform $\vec{S}_i$ by the inverse of $\vec{R}$ from the right side, all correlations are removed (Eq 10). After this operation, all datapoints of $\vec{P}_i$ uncorrelated with white noise.

$$\vec{P}_i = \vec{S}_i \vec{R}^{-1} \qquad (10)$$

The Choleksy decomposition (Eq 8,9), however, requires that the covariance matrix $\vec{C}$ is invertable, that is, of full rank. But this is generally only the case for full bandwith noise. Various other forms of noise, for example narrow-band noise, result in a rank deficiency of the covariance matrix $\vec{C}$. Unfortunately we commonly observe this situation in our data. There exist methods for prewhitening of signals with rank-deficient noise (Doclo and Moonen 2002; Hansen 1998), but this is beyond the scope of this paper. Since all significant covariance values are usually very large, it is technically sufficient to add a very small amount of white noise to the covariance matrix (e.g. with a mean that is only 0.0001% of the covariance values) to make it full rank. While this is theoretically incorrect, it works sufficiently and we have not observed any noticeable differences in the decorrelated data with a rank-deficient prewhitening method and the above method. We are thus using this approach to maximize efficiency.

An alternative approach for whitening is to design a whitening filter and whiten the signal itself before detecting and extracting spikes. This can for example be done by using the matlab function *lpc* to design a filter and use this filter to whiten the signal. This way of processing is less susceptible to the numerical problems mentioned above but is harder to implement in a realtime environment. We used this method of whitening for the results reported in this paper (simulations with exact threshold estimation method).

### *Projection test*

The projection test is entirely calculated on the basis of the prewhitened waveforms $\vec{P}_i$ as described above. In the following, a waveform associated to cluster j is denoted as $\vec{P}_i^{(j)}$ and the center of cluster j is $\vec{P}^{(j)}$.

$$d = \left\| \vec{P}^{(j)} - P^{(k)} \right\| \qquad (11)$$

$$r_i = (\vec{P}_i^{(j)} - P^{(j)}) \bullet \frac{\vec{P}^{(j)} - P^{(k)}}{\left\| \vec{P}^{(j)} - P^{(k)} \right\|} \qquad (12)$$

The distance between two clusters is calculated by taking the norm of the difference between the two centers of cluster j and k (Eq 11). The residual $r_i$ ( scalar) for each spike $\vec{P}_i^{(j)}$ that is assigned to cluster j against cluster k (pairwise comparison between clusters j and k) is calculated by the dotproduct of the difference vector between the center and the spike $\vec{P}_i^{(j)}$, projected onto the vector that connects the two cluster centers (Eq 12).

## *Figures*

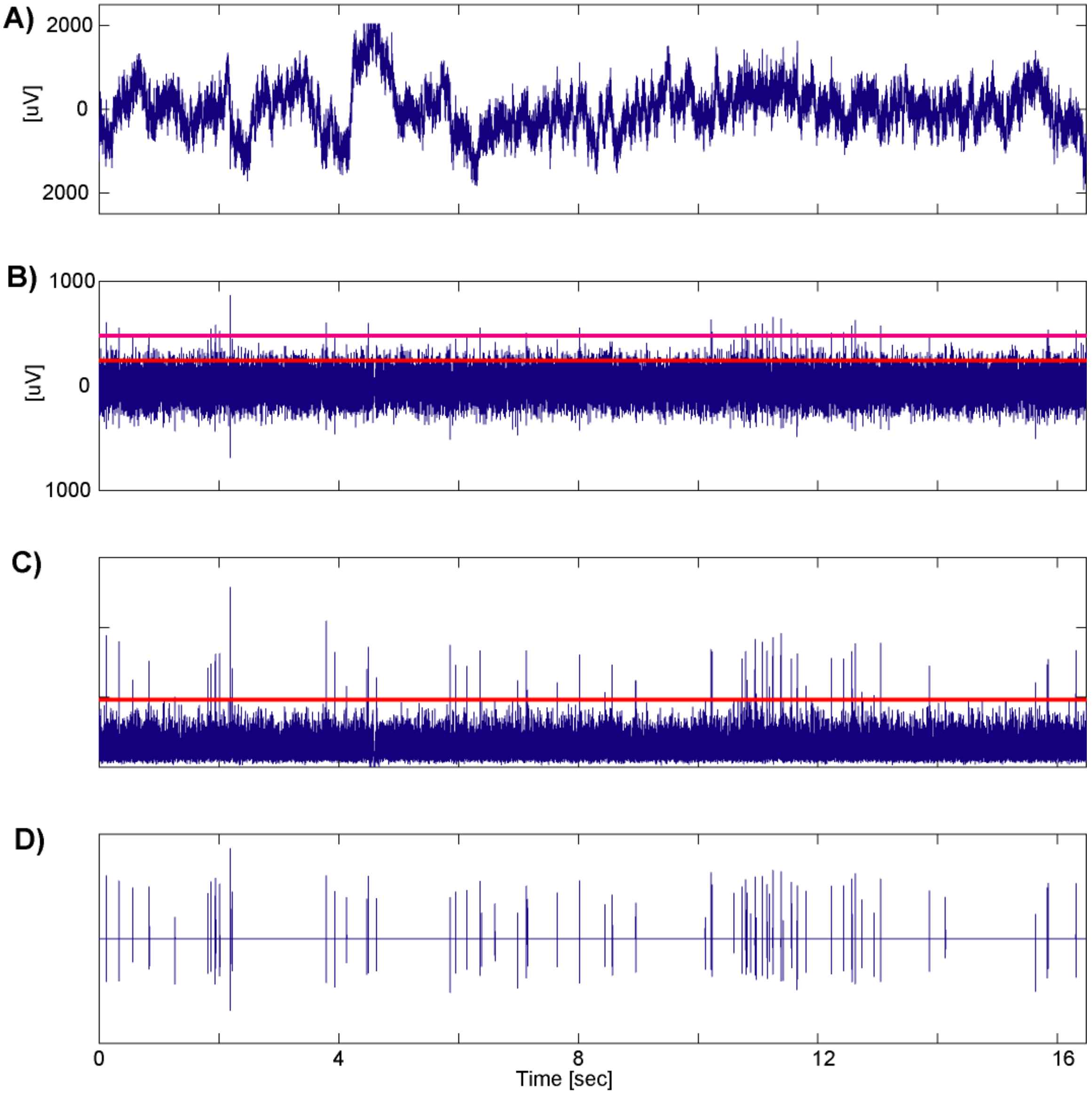

**Figure 1:** Filtering and detection of spikes from continuously acquired data (shown are 412000 timepoints, corresponding to 16.48sec at a sampling rate of 25000Hz). **A)** Raw signal. The amplitude is in units as measured after amplification, not corrected for gain. **B)** Bandpass filtered signal 300-3000Hz. The two lines indicate possible thresholds for direct spike extraction (see text). **C)** Average square root of the power of the signal, calculated with a running window of 1ms and thresholded (line). The y axis is arbitrary. **D)** Position and amplitude of detected spikes (detected in C), but extracted from B).

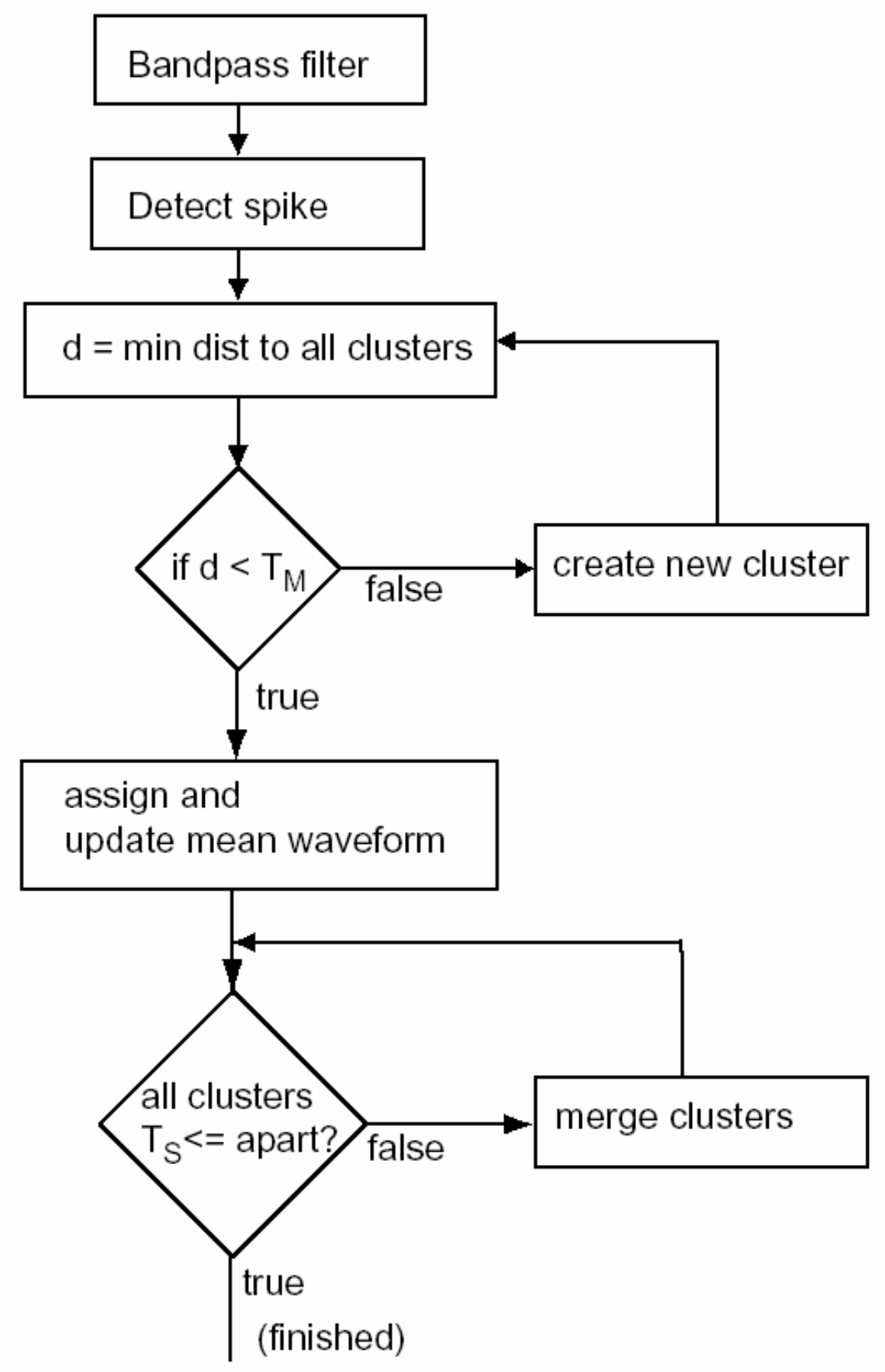

**Figure 2:** Schematic illustration of spike detection and sorting. The signal is (continuously) bandpass filtered 300-3000Hz. Spikes are detected by thresholding a local energy signal that is continuously calculated from the raw filtered signal. After detection and appropriate re-alignment, a distance metric is used to calculate the distance to all known clusters at the current point in time. If the minimal distance is smaller than a threshold $T_M$ , the spike is assigned to this cluster. Otherwise, a new cluster is created and the new spike is assigned to it. The thresholds are automatically and continuously calculated from the noise properties of the raw filtered signal. After assigning a spike to a cluster, that cluster's mean waveform is updated accordingly. This enables tracking of moving electrodes as well as short-term changes due to bursts. After updating the mean waveform, clusters might overlap. If this is the case, they are merged and the spikes assigned to the cluster are reassigned. Periodically, the statistical evaluation criteria (ISI distribution, power spectrum and autocorrelation) as well as the projection test for each pair of clusters are calculated. This allows us to continually discard noise and multi-unit activity.

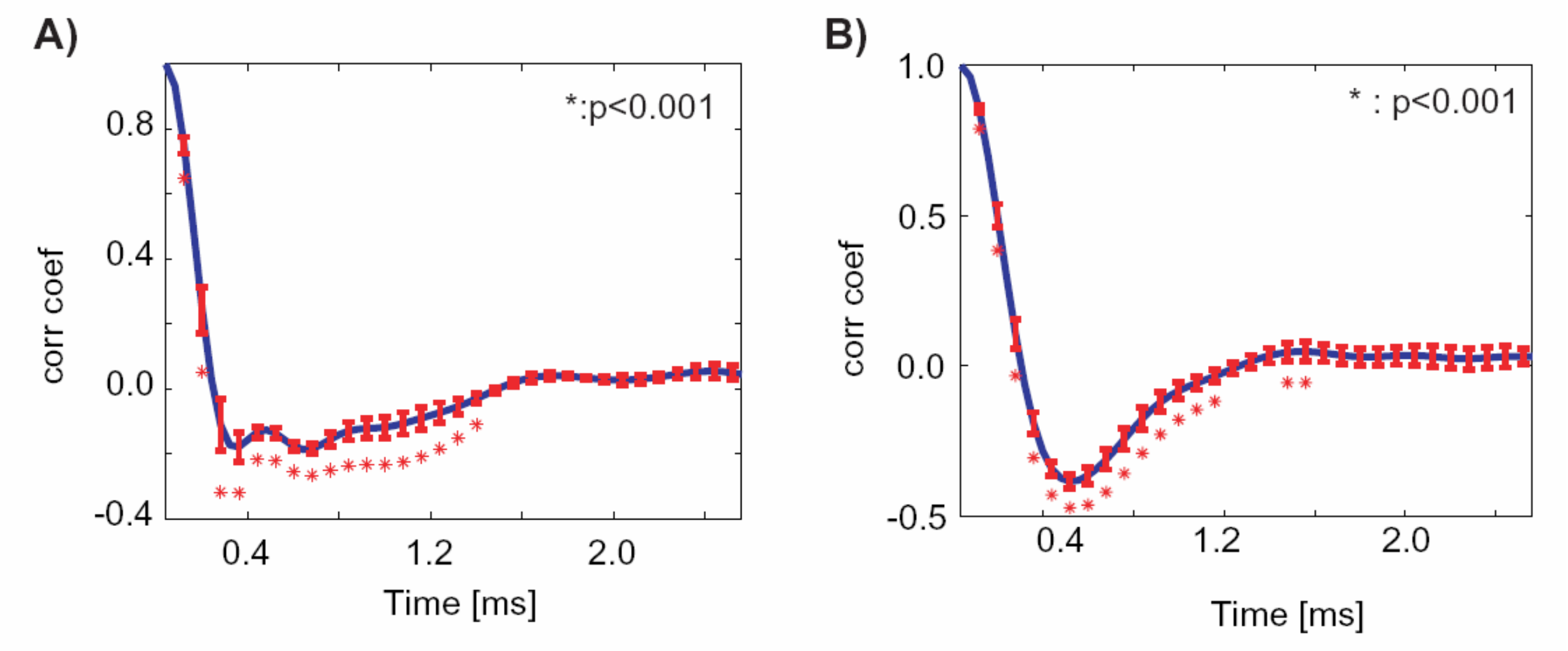

**Figure 3:** Autocorrelation of real (A) and simulated data (B) data. The autocorrelation is calculated from noise traces (which do not contain spikes). **A):** Autocorrelation of the raw signal from real data. Notice that the signal is strongly autocorrelated untill approximately 1.2ms. **(B)** Autocorrelation of simulated data. The autocorrelation remains significant up to 1.2ms (stars indicate p<0.001, t-test for null hypothesis mean = 0). Error bars shown are ± s.d. (n=8542 noise traces).

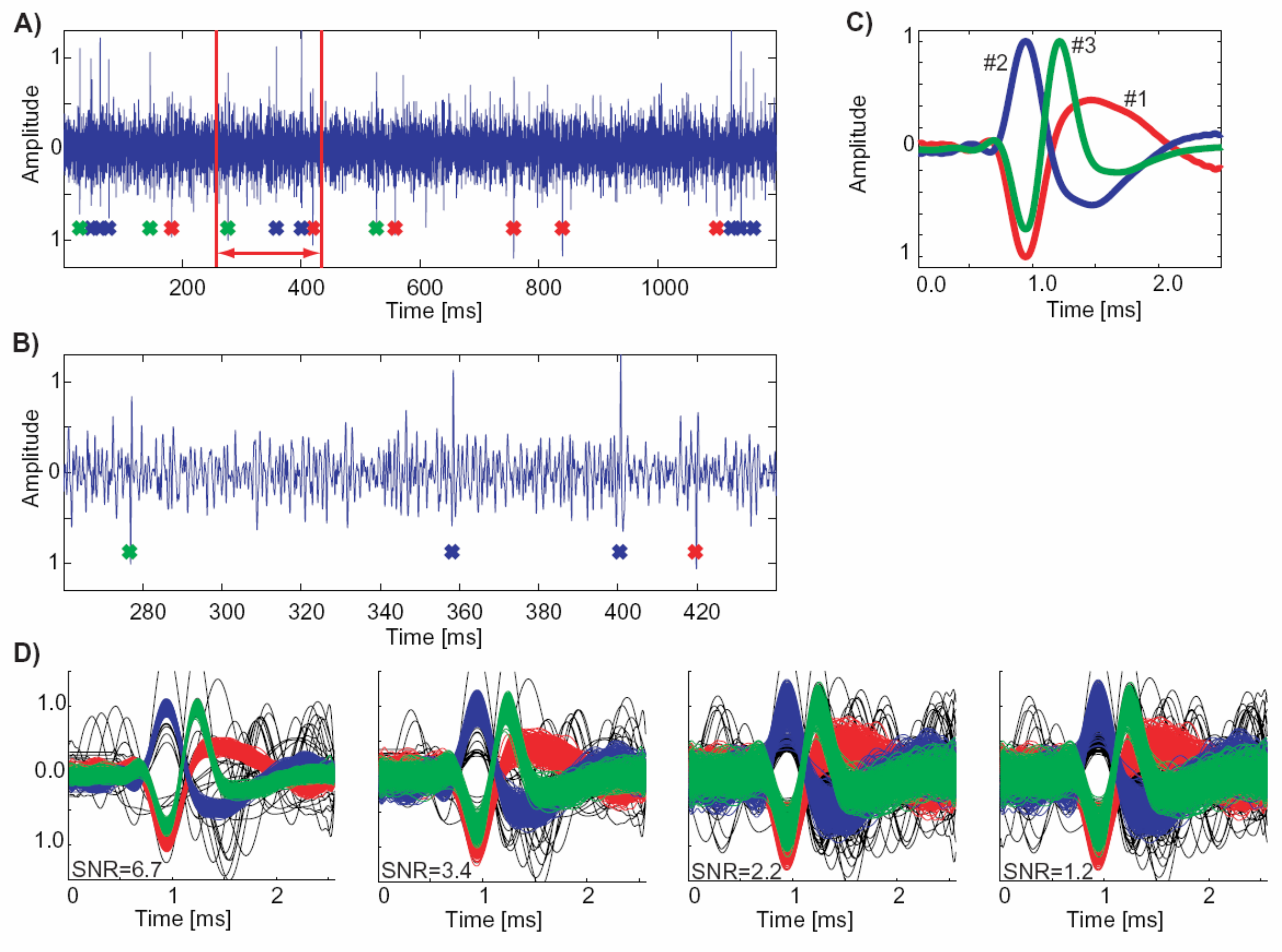

**Figure 4**: Simulated raw signal (dataset 1) from a model extracellular electrode with 3 distinguishable single-units (total length 100s). **A and B)**: Simulated raw signal (bandpass filtered 300-3000Hz) with a noise standard deviation of 0.20 (Level 4 in Table 1). Shown are 1.2s (A) and a zoom-in of 0.3s (B). The colored crosses indicate spikes fired by the randomly firing neurons superimposed on noise. **C)** The mean waveforms of the three single-units. The peak amplitude of each mean waveform is rescaled to 1 (of arbitrary units) to normalize the signal-to-noise ratio. The units fire with a mean frequency of 7, 5 and 4 Hz, respectively (blue, red, green). **D)** Result of detection and sorting for different noise levels (indicated by the respective signal-to-noise (SNR) ratios). The length of the simulated raw data trace was 100s. Correctly sorted spikes are colored (compare to C) while all detected waveforms not associated with any of the 3 units are plotted in black.  (see text for additional discussion).

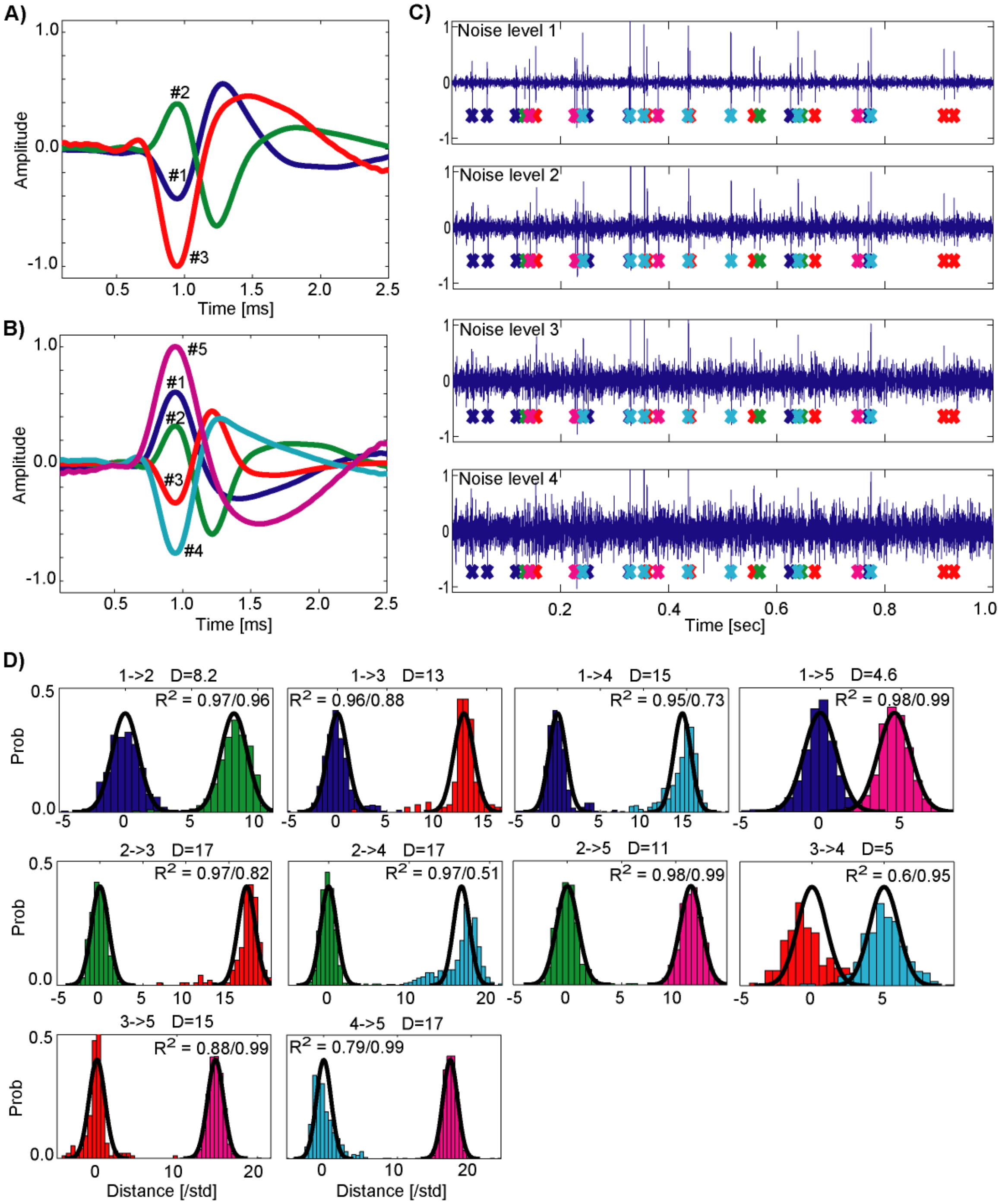

**Figure 5**: Mean waveforms used for simulated dataset 2 (A) and simulated dataset 3 (B). In contrast to dataset 1 (Figure 4), the peak amplitudes of each waveform are scaled randomly, with only one waveform possessing a maximal amplitude of 1. The amplitude is of arbitrary units. **C)**: Raw bandpass filtered data segment of simulated dataset 3 for all 4 levels of noise (from top to bottom). Each segment shown contains spikes of the same 5 neurons. Notice, for example, the two spikes at the right side of the trace (red crosses),

which become hard to detect in noise level 3 and 4. **D)**: Projection test for simulated dataset 3. Shown are all combinations of the 5 neurons shown in B) for noise level 2, matched with color of the histogram and the waveform as well as by number. The histograms depict the probability density function estimated from the residuals of all spikes associated with one cluster. Fit to each distribution is a normal density function with standard deviation = 0. The goodness-of-fit is shown using $R^2$ values. For each combination of neurons, the distance between the two distributions is described by how many standard deviations they are apart (D= in the title of the plots). It can clearly be seen that neurons 1 and 5 as well as 3 and 4 overlap. Also, some of the units are corrupted by noise and thus the $R^2$ value is low. Note that the form of the histogram for the same cluster changes as it is compared to different clusters because the residuals are projected on the line between the two clusters (see text for further discussion).

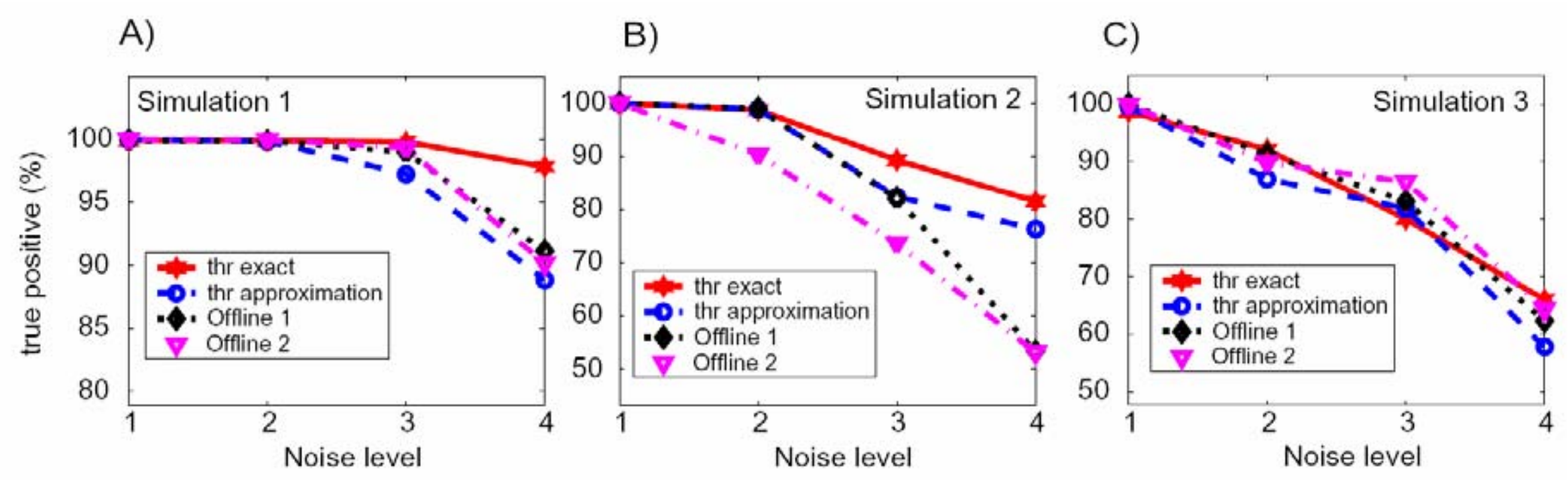

**Figure 6**: Performance comparison. We compared the performance of our algorithm to 2 other offline sorting algorithms (Offline 1 is the Klustakwik Algorithm and 2 the WaveClus Algorithm, see text) , examining the true positives (% of spikes assigned to a given cluster actually belong to the this cluster). For our algorithm we used the two different threshold estimation methods (thr exact and thr approximation). Please see Table 4 for details. The false positive rate is by definition 100-TP.

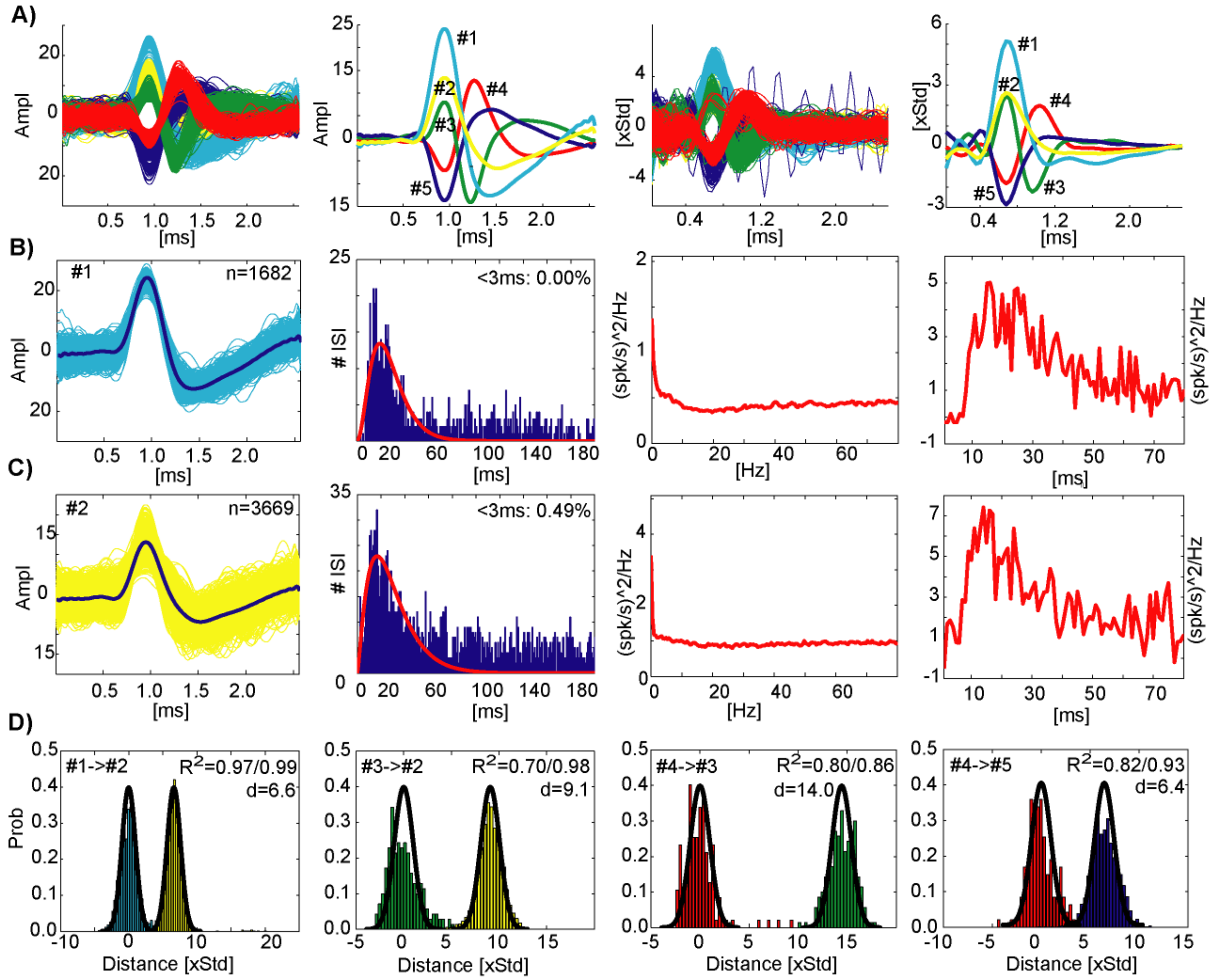

**Figure 7:** Illustration of the sorting process and the tools for evaluation of the sorting result, using real data. All data shown is from the same channel, which was recorded from the right amygdala. Five well-separated neurons could be sorted, with 1682, 3669, 210, 142 and 1534 spikes, respectively (neurons are numbered 1-5 in this order). All subfigures are color matched. **A)**: From left to right, all raw waveforms, mean waveforms, decorrelated raw waveforms and mean decorrelated raw waveforms (see text for discussion of decorrelation). **B and C)**: Details for two of the neurons (#1 and #2, cyan and yellow). From left to right: raw waveforms, ISI histogram, powerspectrum of the ISI and autocorrelation of the ISI. Note that the gamma distribution fitted to the ISI is for illustration purposes only and is not used for evaluation. **D)**: Projection test for the 4 combinations of mean waveforms which are “closest” and could possible overlap/be not well separated. For example, take mean waveforms #1 and #2. They appear to be scaled versions of each other, and clear separation is thus difficult to achieve. It might thus be suspected that they overlap. Consulting the projection test probability density functions shown in the first panel of D), however, allows us to conclude with confidence that these two sets of spikes are well-separated and thus likely represent two unique neurons. The distance (6.6) is big enough and the fit to the theoretical distribution is reasonable.

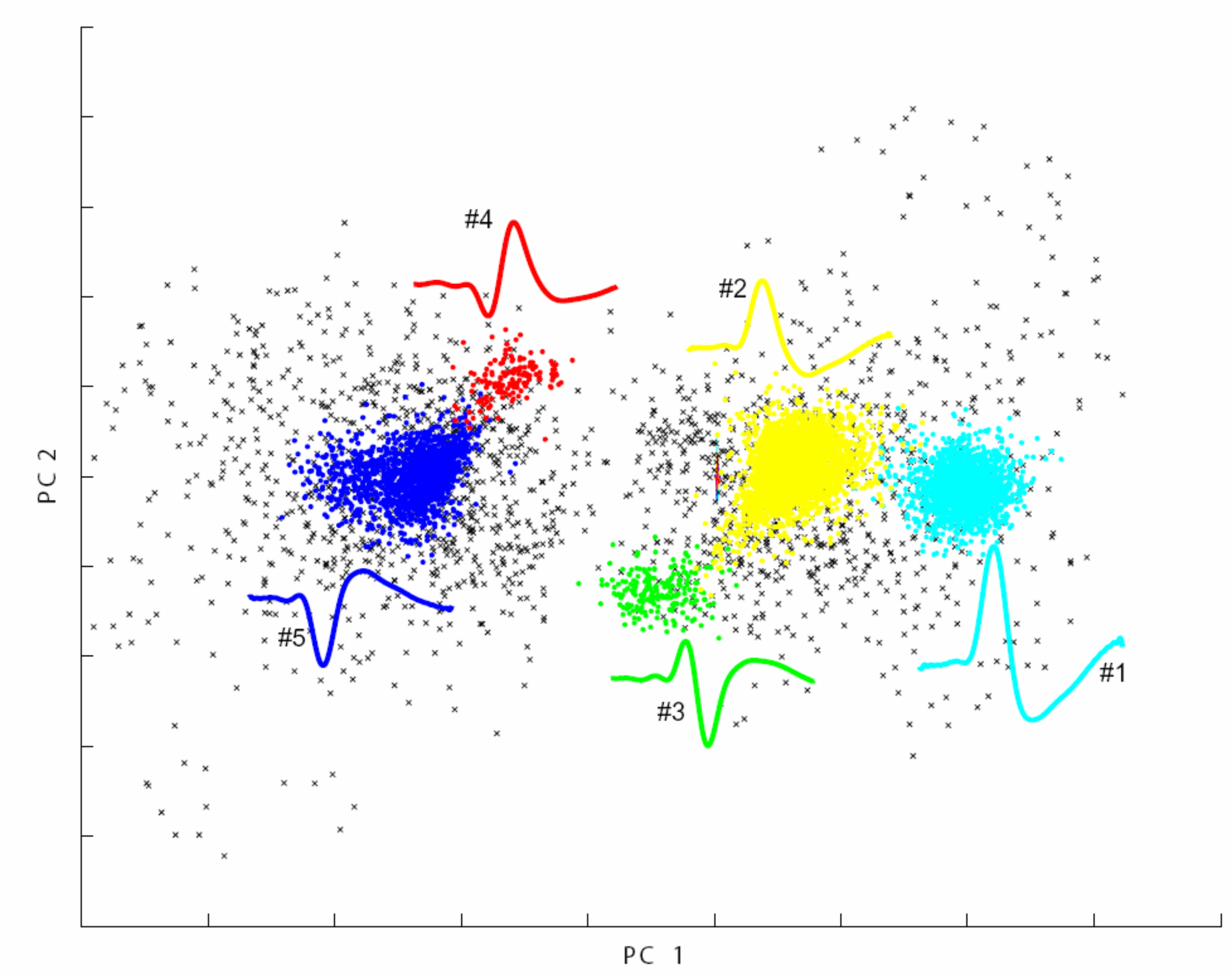

**Figure 8**: Illustration of PC analysis for one channel of real data together with data obtained using our algorithm to sort. Shown is the projection of the first 2 principal components for all waveforms detected on the channel. The colors refer to the same 5 neurons as identified in Figure 7. Black points are detected waveforms which are not assigned to any of the 5 clusters (noise or unsortable). The numbers refer to Figure 7A. This data represents approximately 45 minutes of continuous recording.

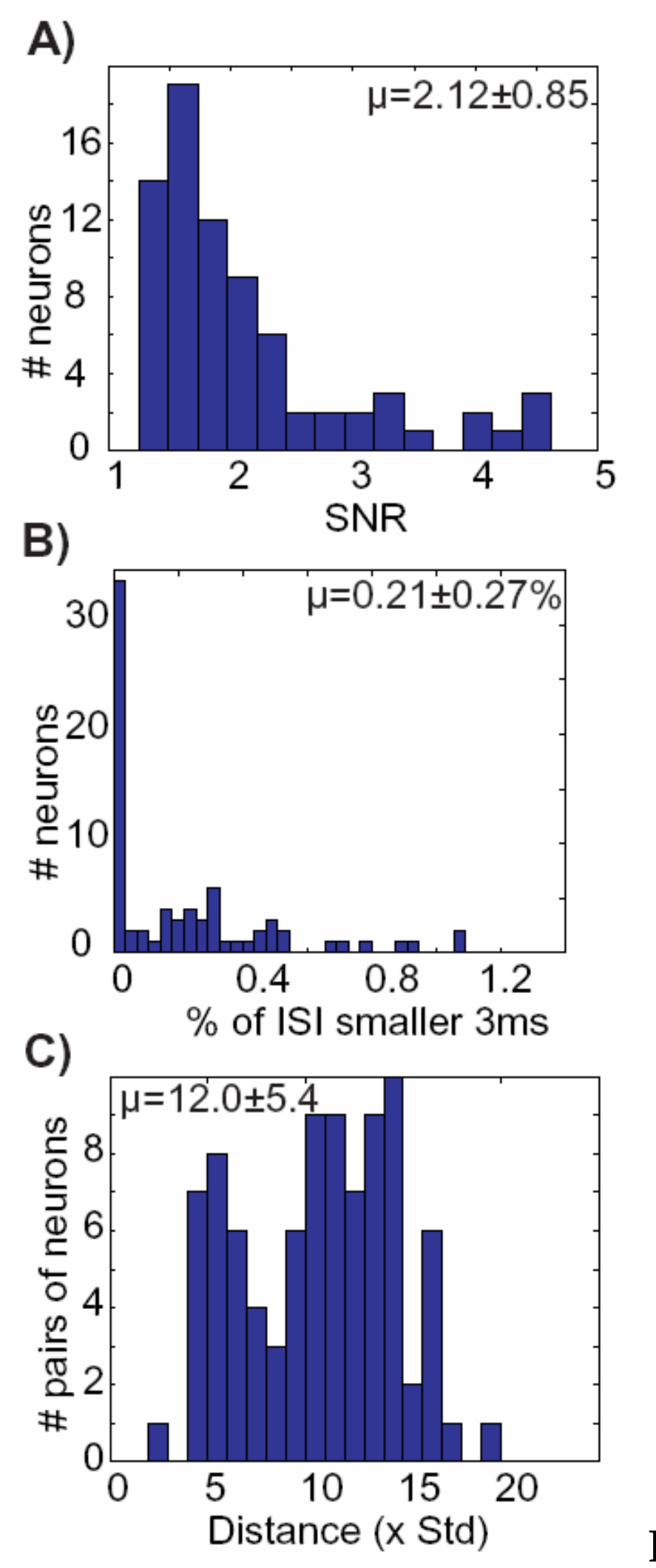

**Figure 9:** Population statistics from the 76 neurons obtained from in vivo recordings which are described in detail in the text. **A):** Histogram of the SNR of all 76 neurons. The SNR is calculated from the mean waveform. The mean SNR was 2.12 ± 0.85 (±s.d.). **B)**: Histogram of the percent of all interspike intervals (ISI) which are shorter than 3ms. The threshold for accepting a neuron is 3%. The mean of all 76 neurons was 0.21 ± 0.27% (± s.d.). **C)**: Histogram of the distance between pair of neurons, calculated using the projection test. This test can only be calculated for channels which have at least one neuron. The mean distance was 12.0 ±5.4 (± s.d.). The distance is expressed as the number of standard deviations of the distribution of waveforms around the mean waveform, which is 1 by design for each neuron.

## ***Tables***

**Table 1**: Simulation 1, consisting of 3 neurons with a peak amplitude of 1 and a firing rate of 5Hz, 7Hz and 4Hz respectively, simulated for 100s. The colors in column 1 refer to figure 4. The 4 noise levels are as follows: 1): s.d = 0.05 and SNR = 6.7 2): s.d. = 0.10 and SNR = 3.4, 3): s.d. = 0.15 and SNR = 2.2, 4): s.d = 0.20 and SNR = 1.2. The case with the lowest SNR is marked bold because it is the situation we most commenly observe in our real data. Abbreviations: Thr: Extraction threshold. TP: True positive, FP: False Positive.

| **N<br>#** | **Spikes<br>#** | **# Detected** [*1]<br>**1 / 2 /3 /4** | | | | **TP** [*2]<br>**1 / 2 /3 /4** | | | | **FP** [*2,3]<br>**1 / 2 /3 /4** | | | | **Misses (Sorting)<br>1 / 2 /3 /4** | | | |
|---|---|---|---|---|---|---|---|---|---|---|---|---|---|---|---|---|---|
| 1<br>*red* | 475 | 475 | 475 | 448 | **366** | 459 | 455 | 414 | **328** | 0 | 0 | 15<br>(12/3) | **54<br>(51/3)** | 16 | 20 | 34 | 38 |
| 2<br>*blue* | 718 | 718 | 718 | 701 | **568** | 693 | 694 | 674 | **521** | 0 | 1<br>(0/1) | 29<br>(7/22) | **101<br>(59/42)** | 25 | 24 | 27 | 47 |
| 3<br>*green* | 383 | 383 | 383 | 377 | **306** | 361 | 354 | 319 | **245** | 0 | 1<br>(0/1) | 2<br>(0/2) | **8<br>(2/6)** | 22 | 29 | 58 | 61 |
| Tot | 1576 | 1576<br>100% | 1576<br>100% | 1526<br>97% | **1240<br>79%** | **1513<br>100%** | **1503<br>100%** | **1407<br>97%** | **1094<br>89%** | 0<br>0% | 2<br>0% | 46<br>3% | **163<br>11%** | 63 | 73 | 119 | 146 |
| Thr | | 4 | 4 | 4 | 4 | | | | | | | | | | | | |

[*1] Percentages for # detected are in terms of % theoretically detectable. [*2] Percentages for TP and FP are in terms of % of all spikes assigned to the sorted cluster. [*3] The numbers in parenthesis represent a split up of the FP into false positives due to noise (first number) and false positives due to assignment to wrong cluster (second number).

**Table 2**: Simulation 2, consisting of 3 neurons with varying amplitude with a firing rate of 5Hz, 7Hz and 4Hz respectively, simulated for 100s. The colors in column 1 refer to figure 5. The 4 noise levels are as follows: 1: s.d. = 0.05 and SNRs of the 3 neurons 4.3, 4.3, 6.9, 2) s.d = 0.10 and SNRs 2.1, 2.1, 3.5, 3): s.d = 0.15 and SNRs 1.4, 1.4, 2.3, 4) s.d. = 0.20 and SNRs 1.1, 1.1, 1.7. The results for the 3rd noise level correspond most closely to what we observe in our data and is marked bold. Abbreviations: Thr: Extraction threshold, TP: True positive, FP: False Positive.

| **N #** | **Spikes #** | **# Detected** [*1] **1 / 2 /3 /4** | | | | **TP** [*2] **1 / 2 /3 /4** | | | | **FP** [*2,3] **1 / 2 /3 /4** | | | | **Misses (Sorting) 1 / 2 /3 /4** | | | |
|---|---|---|---|---|---|---|---|---|---|---|---|---|---|---|---|---|---|
| 1<br>*blue* | 470 | 470 | 466<br>99% | **263**<br>**56%** | 101<br>21% | 442 | 384 | **184** | 25 | 0 | 12<br>(11/1) | **40**<br>**(19/21)** | 4<br>(0/4) | 28 | 82 | **79** | 76 |
| 2<br>*green* | 706 | 706 | 700<br>99% | **395**<br>**56%** | 105<br>15% | 644 | 523 | **235** | 45 | 0 | 1<br>(1/0) | **11**<br>**(8/3)** | 15<br>(12/3) | 62 | 177 | **160** | 60 |
| 3<br>*red* | 392 | 392 | 392<br>100% | **384**<br>**98%** | 274<br>70% | 374 | 344 | **343** | 242 | 0 | 1<br>(0/1) | **151**<br>**(32/119)** | 115<br>(38/77) | 18 | 48 | **41** | 32 |
| Tot | 1568 | 1568<br>100% | 1558<br>99% | **1042**<br>**66%** | 480<br>31% | **1460**<br>**100%** | **1251**<br>**99%** | **762**<br>**82%** | **312**<br>**76%** | 0 | 14<br>1% | **202**<br>**18%** | 134<br>24% | 108 | 307 | **280** | 168 |
| Thr | | 3.0 | 3.0 | 4.0 | 4.5 | | | | | | | | | | | | |

[*1] Percentages for # detected are in terms of % theoretically detectable. [*2] Percentages for TP and FP are in terms of % of all spikes assigned to the sorted cluster. [*3] The numbers in parenthesis represent a split up of the FP into false positives due to noise (first number) and false positives due to assignment to wrong cluster (second number).

**Table 3**: Simulation 3, consisting of 5 neurons with varying amplitude with a firing rate of 5Hz, 7Hz, 4Hz, 6Hz and 9Hz respectively, simulated for 100s. The colors in column 1 refer to figure 5. The 4 noise levels are 1: std = 0.05, SNRs of 5 neurons 4.3, 3.8, 2.8 4.9, 7.9, 2: std=0.10, SNRs 2.1,1.9,1.4,2.4,3.9, 3: std=0.15, SNRs 1.4, 1.3, 0.9, 1.6, 2.6, 4: std=0.20, SNRs 1.1, 0.9, 0.7, 1.2, 1.9. The results for the 3rd noise level correspond closest to what we observe in our data and is marked bold. Notice in noise level 3 that neuron #3 becomes undetectable and in level 4 neurons 1 and 2 merge, which can be seen by the high percentage of false positives in the one remaining cluster. Abbreviations: (m): merged, (-): not detected, Thr: Extraction threshold, * : only detected clusters considered, TP: True positive, FP: False Positive.

| **N<br>#** | **Spikes<br>#** | **# Detected** [*1]<br>**1 / 2 /3 /4** | | | | **TP** [*2]<br>**1 / 2 /3 /4** | | | | **FP** [*2,3]<br>**1 / 2 /3 /4** | | | | **Misses (Sorting)<br>1 / 2 /3 /4** | | | |
|---|---|---|---|---|---|---|---|---|---|---|---|---|---|---|---|---|---|
| 1<br>*blue* | 509 | 508 | 474 | **191** | 112 | 463 | 408 | **0**<br>**(m)** | 0<br>(m) | 6<br>(0/6) | 55<br>(13/42) | **n/a** | n/a | 45 | 66 | **191** | 111 |
| 2<br>*green* | 672 | 671 | 586 | **186** | 100 | 446 | 318 | **65** | 27 | 0 | 10<br>(3/7) | **12**<br>**(2/10)** | 61<br>(28/33) | 225 | 268 | **121** | 73 |
| 3<br>*red* | 375 | 329 | 163 | **31**<br>**8%** | 26 | 296 | 110 | **0**<br>**(-)** | 0<br>(-) | 1<br>(0/1) | 28<br>(4/24) | **n/a** | n/a | 33 | 53 | **31** | 25 |
| 4<br>*l-blue* | 591 | 591 | 590 | **394** | 225 | 539 | 532 | **349** | 182 | 0 | 215<br>(10/205) | **97**<br>**(14/83)** | 128<br>(56/72) | 52 | 58 | **45** | 43 |
| 5<br>*mag.* | 839 | 839 | 839 | **817** | 678 | 787 | 777 | **779** | 611 | 0 | 9<br>(0/9) | **161**<br>**(9/152)** | 119<br>(40/79) | 52 | 62 | **38** | 67 |
| | 2986 | 2938<br>98% | 2652<br>89% | **1619**<br>**54%** | 1141<br>38% | 2531<br>100% | 2145<br>87% | **1193**<br>**82%** | 820<br>58% | 7 | 317<br>13% | **270**<br>**18%** | 308<br>42% | 407 | 507 | **426** | 319 |
| Thr | | 3 | 3 | 4 | 4 | | | | | | | | | | | | |

[*1] Percentages for # detected are in terms of % theoretically detectable. [*2] Percentages for TP and FP are in terms of % of all spikes assigned to the sorted cluster.

[*3] The numbers in parenthesis represent a split up of the FP into false positives due to noise (first number) and false positives due to assignment to wrong cluster (second number).

| | Noise Level | Percentage of assigned spikes which are TP (100-x is FP) | | | | Nr valid clusters found | | | | Percentage of spikes missed | | | |
|---|---|---|---|---|---|---|---|---|---|---|---|---|---|
| | | **Approx** | **Exact** | **Offline 1** | **Offline 2** | **Approx** | **Exact** | **Offline 1** | **Offline 2** | **Approx** | **Exact** | **Offline 1** | **Offline 2** |
| **Simulation 1** | 1 | 100.00 | 99.91 | 99.91 | 100.00 | 3 | 3 | 3 | 3 | 4.00 | 3.87 | 3.68 | 2.92 |
| | 2 | 99.86 | 99.91 | 99.86 | 99.95 | 3 | 3 | 3 | 3 | 4.63 | 5.84 | 3.49 | 2.92 |
| | 3 | 97.25 | 99.80 | 98.98 | 99.33 | 3 | 3 | 3 | 3 | 7.80 | 9.32 | 6.16 | 8.00 |
| | 4 | 88.82 | 97.84 | 91.12 | 90.14 | 3 | 3 | 3 | 3 | 11.77 | 14.97 | 11.05 | 17.34 |
| | **mean** | **96.48** | **99.36** | **97.47** | **97.36** | **3** | **3** | **3** | **3** | **7.05** | **8.50** | **6.09** | **7.79** |
| **Simulation 2** | 1 | 100.00 | 100.00 | 100.00 | 100.00 | 3 | 3 | 3 | 3 | 6.89 | 3.76 | 6.51 | 6.57 |
| | 2 | 98.72 | 98.83 | 98.97 | 90.37 | 3 | 3 | 3 | 3 | 19.70 | 6.51 | 16.05 | 18.29 |
| | 3 | 82.37 | 89.28 | 82.06 | 73.60 | 3 | 3 | 3 | 2 | 26.87 | 25.97 | 23.42 | 45.87 |
| | 4 | 76.33 | 81.53 | 53.30 | 53.10 | 3 | 3 | 2 | 2 | 35.00 | 24.48 | 35.41 | 35.21 |
| | **mean** | **89.36** | **92.41** | **83.58** | **79.26** | **3** | **3** | **2.75** | **2.5** | **22.12** | **15.18** | **20.35** | **26.49** |
| **Simulation 3** | 1 | 99.68 | 98.74 | 99.96 | 99.92 | 5 | 5 | 5 | 5 | 13.85 | 10.09 | 14.06 | 11.61 |
| | 2 | 86.97 | 92.08 | 91.23 | 89.83 | 5 | 5 | 5 | 4 | 19.12 | 19.02 | 18.06 | 62.63 |
| | 3 | 81.84 | 80.01 | 83.11 | 86.50 | 3 | 4 | 3 | 3 | 26.31 | 31.46 | 24.77 | 34.03 |
| | 4 | 57.70 | 65.96 | 62.26 | 64.26 | 3 | 4 | 3 | 3 | 7.13 | 41.61 | 21.74 | 27.86 |
| | **mean** | **81.55** | **84.20** | **84.14** | **85.12** | **4** | **4.5** | **4** | **3.75** | **16.60** | **25.54** | **19.65** | **34.03** |

**Table 4**: Comparison of sorting results for the two different threshold estimation methods (Columns Approximation and Exact) as well as two other algorithms (Columns Offline 1 and 2, see text). Percentages of true positives (TP) are specified in terms of percent of all spikes assigned to the cluster. False positives (FP) are thus by definition 100-TP. The column “nr valid clusters found” specifies how many of the original clusters were found. The right column “percentage of spikes missed” specifies what percentage of all correctly detected spikes (spikes which are known to belong to one of the simulated neurons, excluding noise detections) were not assigned to the correct cluster. This number includes both spikes assigned to background noise or the wrong cluster.